\documentclass[prodmode,acmcsur]{acmsmall}
\input{Qcircuit.tex}
\renewcommand{\Qcircuit}[1][0em]{\xymatrix @*=<#1>}
\usepackage{color,multirow,subfigure}
\usepackage{graphicx}
\usepackage[full]{complexity}

\begin{document}
\newcommand{\controlu}{*-=[][F]{\phantom{\bullet}}}
\newcommand{\multistate}[2]{*+{\hphantom{#2}} \POS[0,0].[#1,0] !C
  *{#2} \POS[0,0].[#1,0] \drop\frm{}}
\newcommand{\ghoststate}[1]{*+{\hphantom{#1}} }
\newcommand{\ccteq}[1]{\multistate{#1}{=}}
\newcommand{\ccteqg}{\ghoststate{=}}
\newcommand{\csl}{{\ \;
    \backslash }}

\markboth{M. Saeedi and I. L. Markov}{Synthesis and Optimization of Reversible Circuits - A Survey}

\title{\large{Synthesis and Optimization of Reversible Circuits - A Survey}}

\author{MEHDI SAEEDI
\affil{Amirkabir University of Technology}
IGOR L. MARKOV
\affil{University of Michigan}
}

\begin{abstract}
Reversible logic circuits have been historically motivated by theoretical research in low-power electronics as well as practical improvement of bit-manipulation transforms in cryptography and computer graphics. Recently, reversible circuits have attracted interest as components of quantum algorithms, as well as in photonic and nano-computing technologies where some switching devices offer no signal gain. Research in generating reversible logic distinguishes between circuit synthesis, post-synthesis optimization, and technology mapping. In this survey, we review algorithmic paradigms --- search-based, cycle-based, transformation-based, and BDD-based --- as well as specific algorithms for reversible synthesis, both exact and heuristic. We conclude the survey by outlining key open challenges in synthesis of reversible and quantum logic, as well as most common misconceptions.
\end{abstract}

\begin{bottomstuff}
M. Saeedi is currently with the Department of Electrical Engineering, University of Southern California.\\
Authors' address: M. Saeedi, Department of Electrical Engineering, University of
Southern California, Los Angeles, CA 90089-2562; email: msaeedi@usc.edu;
I. L. Markov, Department of EECS, University of Michigan, Ann Arbor, MI 48109; email: imarkov@eecs.umich.edu.
\end{bottomstuff}

\maketitle

\section{Introduction}
A computation is \emph{reversible} if it can be `undone' in the sense that the output contains sufficient information to reconstruct the input, i.e., no input information is erased \cite{Toffoli80}. It is also common to require that no information is duplicated. In Computer Science, reversible transformations have been popularized by the Rubik's cube and sliding-tile puzzles, which fueled the development of new algorithms, such as iterative-deepening A*-search \cite{Korf99}. Prior to that, reversible computing was proposed to minimize energy loss due to the erasure and duplication of information. Today, reversible information processing draws motivation from several sources.

\begin{itemize}
\item [$\bullet$] \textbf{Considerations of power consumption} prompted research on reversible computation, historically. In 1949, John Von Neumann estimated the minimum possible energy dissipation per bit as $k_BT \ln 2$ where $k_B=1.38065 \times 10^{-23} \mathrm{J}/\mathrm{K}$ is the Boltzmann constant and $T$ is the temperature of environment \cite{Neumann66}. Subsequently, \citeN{Landauer61} pointed out that the irreversible erasure of a bit of information consumes power and dissipates heat. While reversible designs avoid this aspect of power dissipation, most power consumed by modern circuits is unrelated to computation but is due to clock networks, power and ground networks, wires, repeaters, and memory. A recent trend in low-power electronics is to replace logic reversibility by charge recovery, e.g., through dual-rail encoding where the $01$ combination represents a logical $0$ and $10$ represents a logical $1$ \cite{Kim:2005}.\footnote{While charge recovery reminds \emph{conservative logic} \cite{Fredkin82}, its essential property is to avoid dissipating electric charges by exchanging them. This property requires transistor-level support and is not specific to logic circuits as it also applies to clock networks.}
\vspace{1mm}
\item [$\bullet$] \textbf{Signal processing, cryptography, and computer graphics} often require reversible transforms, where all of the information encoded in the input must be preserved in the output. A common example is swapping two values $a$ and $b$ without intermediate storage by using bitwise XOR operations $a=a \oplus b$, $b=a \oplus b$, $a=a \oplus b$. Given that reversible transformations appear in bottlenecks of commonly-used algorithms, new instructions have been added to the instruction sets of various microprocessors such as \texttt{vperm} in PowerPC AltiVec, \texttt{bshuffle} in Sun SPARC VIS, \texttt{permute} and \texttt{mix} in HP PA-RISC, \texttt{pshufb} in Intel IA32 and \texttt{mux} in Intel IA64 to improve their performance \cite{McGregor03}. In particular, the performance of cryptographic algorithms DES, Twofish and Serpent, as well as string reversals and matrix transpositions, can be considerably improved by the addition of bit-permutation instructions \cite{Shi:2000,Hilewitz08}. In another example, the reversible \emph{butterfly} operation is a key element for Fast Fourier Transform (FFT) algorithms and has been used in application-specific Xtensa processors from Tensilica. Reversible computations in these applications are usually short and hand-optimized.

\vspace{1mm}
\item [$\bullet$] \textbf{Program inversion and reversible debugging} generalize the `undo' feature in integrated debugging environments and allow reconstructing sequences of decisions that lead to a particular outcome. Automatic program inversion \cite{Gluck:2005} and reversible programming languages \cite{Yokoyama:2008,DeVosBook} allow reversible execution. Reversible debugging \cite{Visan09} supports reverse expression watch-pointing to provide further examination of a problematic event.

\vspace{1mm}
\item [$\bullet$] \textbf{Networks on chip} with mesh-based and hypercubic topologies \cite{Dally:2003} perform \emph{permutation routing} among nodes when each node can both send and receive messages. To route a message, regular permutation patterns such as bit-reversal, complement and transpose are applied to minimize the number of communication steps.

\vspace{1mm}
\item [$\bullet$] \textbf{Nano- and photonic circuits} \cite{Politi09,Gao2010} are made up of devices without gain, and they cannot freely duplicate bits because that requires energy. They also tend to recycle available bits to conserve energy. Generally, building nano-size switching devices with gain is difficult because this requires an energy distribution network. Therefore, reversibility is fundamentally important to nano-scale computing, although specific constraints may vary for different technologies.

\vspace{1mm}
\item [$\bullet$] \textbf{Quantum computation} \cite{Nielsen00} is another motivation to study reversible computation because unitary transformations in quantum mechanics are reversible. Quantum algorithms have been designed to solve several problems in polynomial time \cite{Bacon:2010,Childs10}, where best-known conventional algorithms take more than polynomial time.\footnote{\BQP\,(Bounded-Error Quantum Polynomial-Time) is the class of problems solvable by a quantum algorithm in polynomial time with at most $\frac{1}{3}$ probability of error. \P$\,$ is the class of problems solvable by a deterministic Turing machine in polynomial time. Quantum computers have attracted attention as several \BQP\, problems of practical interest are expected to be outside \P.} A key example is number-factoring, which is relevant to cryptography. While unitary transformations can be difficult to work with in general, many prominent quantum algorithms contain large blocks with reversible circuits that do not invoke the full power of quantum computation, e.g., for arithmetic operations \cite{Beckman96,Meter05,TakahashiQIC08,MarkovQIC2012}. Circuits for quantum error-correction contain large sections of reversible circuits that implement GF(2)-linear transformations \cite{Aaronson04}.

\end{itemize}

\begin{figure}[tb]
\centering
\includegraphics[height=40mm]{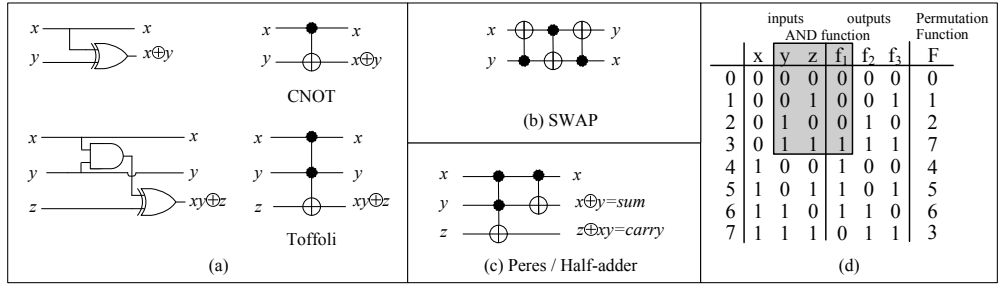}
\caption{Expressing CNOT and Toffoli gates using AND and XOR gates (a), swapping two values $x$ and $y$ by three XOR operations (CNOT) as a reversible circuit (b), a reversible half-adder circuit (c), a sample reversible function (d).}
\label{fig:Fig1}
\end{figure}

In software and hardware applications of reversible information processing, sequences of reversible operations can be viewed as reversible circuits. For example, swapping two values $x$ and $y$ with a sequence of three XOR or CNOT gates (shown in Fig. \ref{fig:Fig1}a), operations $x=x\oplus y$, $y=x\oplus y$, and $x=x\oplus y$ is illustrated in Fig. \ref{fig:Fig1}b by a circuit. Such circuits are particularly useful in quantum computing. Reversibility prohibits loops and explicit fanouts in circuits,\footnote{Read-only fanouts do not conflict with this requirement as illustrated by line $x$ in Fig. \ref{fig:Fig1}c, and arbitrary fanouts can be simulated using ancilla lines, as we show in Fig. \ref{fig:copy}b.} and each gate must have an equal number of inputs and outputs with unique input-to-output assignments. Such peculiar features of reversible circuits prevent the use of existing algorithms and tools for circuit synthesis and optimization. \emph{Reversible logic synthesis} is the process of generating a compact reversible circuit from a given specification. Research on reversible logic synthesis has attracted much attention after the discovery of powerful quantum algorithms in the mid 1990s \cite{Nielsen00}. Closely related techniques have also been motivated by other applications, e.g., the decomposition of permutations into tensor products is an important step in deriving fast algorithms and circuits for digital signal processing (Fourier and cosine transforms, etc.) \cite{Egner:1997}.

This survey discusses methodologies, algorithms, benchmarks, tools, open problems and future trends related to the synthesis of combinational reversible circuits.
The remaining part is organized as follows. In Section \ref{sec:basic} basic concepts are introduced. We outline the process of reversible synthesis in Section \ref{sec:picture}, including optimization and technology mapping. Algorithmic details are examined in Sections \ref{sec:syn} and \ref{sec:post_syn}. Available benchmarks and tools for reversible logic are introduced in Section \ref{sec:bnch}. Finally, we discuss open challenges in reversible circuit synthesis in Section \ref{sec:open}.

\section{Basic Concepts} \label {sec:basic}
In this section, we introduce reversible logic gates and quantum gates, as well as reversible and quantum circuits. Representations of reversible functions and cost models for reversible gates are also discussed.

\subsection{Reversible Gates and Circuits}
Let $A$ be a finite set and $f: A \rightarrow A$ a one-to-one and onto (bijective) function, i.e., a permutation. For instance, the function $g=(1,5,3,2,0,4,6,7)$ is a permutation over $\{0,1,\cdots,7\}$ where $g(0)=1$, $g(1)=5$, $g(2)=3$, etc. The set of all permutations on $A=\{0,1,\cdots,2^n-1\}$ forms the \emph{symmetric group}\footnote{In abstract algebra, a \emph{group} is a set with a binary operation on it, viewed as multiplication, which is associative $(a\cdot b)\cdot c=a\cdot(b\cdot c)$, has a neutral element $e$ such that $a\cdot e=e\cdot a= a$, and admits an inverse for every element $a\cdot a^{-1}=e$.} $S_n$ on $A$. A \emph{reversible Boolean function} is a multi-output Boolean function with as many outputs as inputs, that is reversible. Fig. \ref{fig:Fig1}d illustrates a reversible function on three variables that implements the permutation $F=\{0,1,2,7,4,5,6,3\}$.

\vspace{1mm}
\textbf{Cycles}. A \emph{cycle} $(a_1, a_2, \cdots, a_k)$ is a permutation such that $f(a_1)=a_2$, $f(a_2)=a_3$, ..., and $f(a_k)=a_1$. For example, $g$ can be written as $(0,1,5,4)(2,3)(6)(7)$. The \emph{length} of a cycle is the number of elements it contains. A cycle of length two is called a \emph{transposition}. A cycle of length $k$ is called a \emph{$k$-cycle}. 1-cycles, e.g., (6) and (7) in $g$, are usually omitted. Cycles $c_1$ and $c_2$ are \emph{disjoint} if they have no common members. Any permutation can be written as a product of disjoint cycles. This decomposition is unique up to the order of cycles. The composition of two disjoint cycles does not depend on the order in which the cycles are applied --- disjoint cycles \emph{commute}. In addition, a cycle may be written in different ways as a product of transpositions, e.g., $g=(0,1) (0,5) (0,4) (2,3)$ and $g=(4,5)(0,1)(1,5)(4,5)(0,4)(2,3)$. A cycle is \emph{even} (\emph{odd}) if it can be written as an even (odd) number of transpositions, i.e., a $k$-cycle is odd (even) if $k$ is even (odd). The same definition applies to even and odd permutations in general.

\begin{figure}[tb]
\centering
\includegraphics[height=21mm]{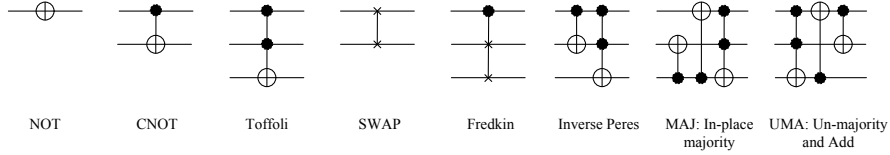}
	\caption {Basic reversible gates. The Peres gate (reversible half-adder) is defined in Fig. \ref{fig:Fig1}c. The MAJ and UMA gates together
   form a full-adder gate, used in [Cuccaro et al. 2005] to build reversible multi-bit adders.
} \label{fig:Rgates}
\end{figure}

\vspace{1mm}
\textbf{Reversible gates}. A \emph{reversible gate} realizes a reversible function. For a gate $g$, the gate $g^{-1}$ implements the inverse transformation. Common reversible gates are illustrated in Fig. \ref{fig:Rgates}.
\begin{itemize}
\item [$\bullet$] A \emph{multiple-control Toffoli gate} \cite{Toffoli80} C$^m$NOT$(x_1, x_2, \cdots, x_{m+1})$ passes the first $m$ lines, \emph{control} lines, unchanged. This gate flips the $(m+1)$-th line, \emph{target} line, if and only if each positive (negative) control line carries the 1 (0) value. 
    For $m=0,1,2$ the gates are named NOT (N), CNOT (C), and Toffoli (T), respectively. These three gates compose the universal NCT library.
\item [$\bullet$] A {\em multiple-control Fredkin gate} \cite{Fredkin82} Fred$(x_1, x_2, \cdots,x_{m+2})$ has two target lines $x_{m+1},x_{m+2}$ and $m$ control lines $x_1, x_2, \cdots,x_m$. The gate interchanges the values of the targets if the conjunction of all $m$ positive (negative) controls evaluates to 1 (0). 
    For $m=0,1$ the gates are called SWAP (S) and Fredkin (F), respectively.
\item [$\bullet$] A {\em Peres gate} \cite{Peres85} $P(x_1,x_2, x_3)$ has one control line $x_1$ and two target lines $x_2$ and $x_3$. It represents a C$^2$NOT($x_1,x_2, x_3$) and a CNOT($x_1, x_2$) in a cascade.
\item [$\bullet$] An \emph{in-place majority} (MAJ) gate computes the majority of three bits in place \cite{Cuccaro:quant-ph0410184}, and provides the carry bit for addition. Cascading it with an \emph{Un-majority and Add} (UMA) gate \cite{Cuccaro:quant-ph0410184} forms a full adder.
\end{itemize}

In multiple-control Toffoli and Fredkin gates, each line is either control or target. The order of controls is immaterial and so is the order of targets, but interchanging controls with targets will create a different gate. A multiple-control Toffoli (or Fredkin) gate implements a single transposition if only incident bit-lines are considered. The transposition is determined by the controls of the gate. If an extended set of bit-lines is considered, these gates will implement sets of disjoint transpositions.\footnote{If logic 0 and 1 are encoded as 01 and 10, respectively (dual-rail), SWAP performs inversion and the Fredkin gate models the function of the CNOT gate.} Multiple-control Toffoli and Fredkin gates are self-inverse. For a Peres gate $P(x_1,x_2, x_3)$, the inverse Peres is the CNOT($x_1, x_2$) C$^2$NOT($x_1,x_2, x_3$) pair. For MAJ and UMA gates, the inverse gates can be constructed by reordering the CNOT and Toffoli gates.

\vspace{1mm}
\textbf{Reversible circuits}. A combinational \emph{reversible circuit} is an acyclic combinational logic circuit in which all gates are reversible, and are interconnected without explicit fanouts and loops. In this survey, gates in a circuit diagram are processed from left to right. A reversible half-adder circuit in Fig. \ref{fig:Fig1}c implements the conventional half-adder when $z=0$.\footnote{Similar to the conventional arithmetic circuits that are typically designed in terms of half- and full-adders, identifying useful blocks such as half-adders is also common in reversible logic \cite{Beckman96,Meter05,Cuccaro:quant-ph0410184,MarkovQIC2012}.} For a set of gates $g_1$, $g_2$, ..., $g_k$ cascaded in a circuit $C$ in sequence, the circuit $C^{-1} = g_k^{-1} g_{k-1}^{-1} \cdots g_1^{-1}$ (where $g_i^{-1}$ is the inverse of $g_i$) implements the inverse transformation with respect to $C$. Different circuits computing the same function are considered \emph{equivalent}. For example, circuits $C_1$=SWAP($x,y$) and $C_2=$CNOT($x,y$) CNOT($y,x$) CNOT($x,y$) (Fig. \ref{fig:Fig1}b) are equivalent. For a library $\mathcal{L}$, an \emph{$\mathcal{L}$-circuit} is composed only of gates from $\mathcal{L}$. A permutation is \emph{$\mathcal{L}$-constructible} if it is computable by an $\mathcal{L}$-circuit. When the library consists of a single gate (type), we use the gate name instead of $\mathcal{L}$. We call permutations implementable with only NOT, CNOT, or Toffoli gates N-\emph{constructible}, C-\emph{constructible}, or T-\emph{constructible}, respectively. $S_{n}$ has $2^n$ N-constructible, $\prod_{i=0}^{n-1}(2^n-2^i)$ C-constructible, and $(1/2)(2^n-n-1)!$ T-constructible permutations \cite{ShendeTCAD03}. Every even permutation is NCT-constructible \cite{DeVos2002,ShendeTCAD03}. When dealing with $n$ bits, reversible logic synthesis searches for solutions in a space of $O(n2^n)$ elements \cite{SaeediJETC10}. A function $f$ is {\em affine-linear}, or {\em linear} in short, if $f(x_1 \oplus x_2) = f (x_1) \oplus f (x_2)$ where $\oplus$ is a multi-bit XOR operation. NC-constructible permutations are linear functions and vice versa \cite{PatelQIC08}. NCTSFP is the library consisting of NCT gates with SWAP, Fredkin and Peres gates added.

\begin{figure}[tb]
\centering
\includegraphics[height=22mm]{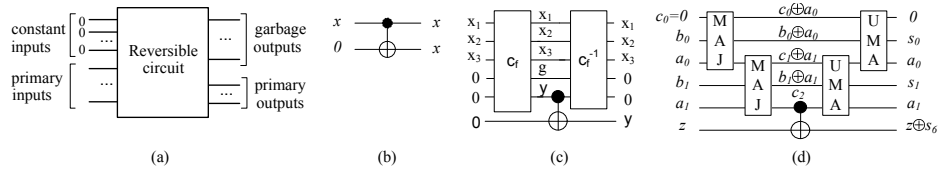}
\caption{The structure of inputs and outputs in a reversible circuit (a), explicit fanout in a reversible circuit with a CNOT gate and one ancilla (b), a reversible circuit for computing $y = f(x_1, x_2, x_3)$ described by $C_f$ and garbage line $g$ (c), a 2-bit ripple-carry adder [Cuccaro et al. 2005] (d).}
\label{fig:copy}
\end{figure}

\vspace{1mm}
\textbf{Ancilla lines}. There are $2^n!$ distinct reversible functions on $n$ variables which are permutations for $2^n$ elements. However, $\sum_{i=1}^{n}{(2^i)}^{2^n}\simeq2^{n2^n}$ irreversible multiple-output (from 1 to $n$) functions exist.
To make the specification reversible, input/output should be added. The added lines are called \emph{ancillae} and typically start out with the 0 or 1 constant. An ancilla line whose value is not reset to a constant at the end of the computation is called a \emph{garbage} line. Unconstrained outputs of ancillae lines in the truth table are called \emph{don't cares} (DC). For an irreversible specification where each output combination can be repeated up to $M$ times, $g = \lceil \log_2 M \rceil$ ancillae are required to build a reversible circuit \cite{MaslovTCAD04}.
For example, at least two garbage lines ($f_2$ and $f_3$) and one constant line ($x$) are required to make the AND gate reversible as shown in Fig. \ref{fig:Fig1}d. Every odd permutation can be implemented with an NCT-circuit using one ancilla bit \cite{ShendeTCAD03}. The Toffoli gate can be used with one constant line to compute the NAND function, i.e., C$^2$NOT($a,b,1$), making Toffoli a universal gate in the Boolean domain. In general, the number of constant lines plus primary inputs is equal to the number of garbage lines plus primary outputs. See Fig. \ref{fig:copy}a for an illustration. A reversible copying gate or explicit fanout can be simulated by a CNOT and one ancilla, which leaves no garbage bit at the output as illustrated in Fig. \ref{fig:copy}b.

\vspace{1mm}
\textbf{Reversible implementations}. \citeN{Toffoli80} proposed a generic NCT-circuit construction for an arbitrary reversible or irreversible function. For an implementation of any irreversible function $f(x)$, its reversible implementation can be described in the form $(x, y)  \mapsto (x, y \oplus f(x))$. This specification is reversible since composing it with itself produces $(x, y \oplus f(x) \oplus f(x))=(x,y)$. Given a conventional circuit for $f$, a reversible circuit can be constructed by making each gate reversible using a set of \emph{temporary} lines if necessary. To reuse these temporary lines again, their values should be restored to their initial values. To restore the values, first copy function outputs to a set of ancillae with initial values 0 and then run the obtained reversible circuit in reverse to recover the starting values \cite{Bennett73} as illustrated in Fig. \ref{fig:copy}c. Fig. \ref{fig:copy}d shows a 2-bit ripple-carry adder with one ancilla \cite{Cuccaro:quant-ph0410184} where values of $a_0$ and $a_1$ are recovered after computation. Note that if the values of temporary lines in a circuit are not restored, this circuit cannot be inverted and is not convenient as building blocks for larger circuits.

\begin{figure}[tb]
\centering
\includegraphics[height=55mm]{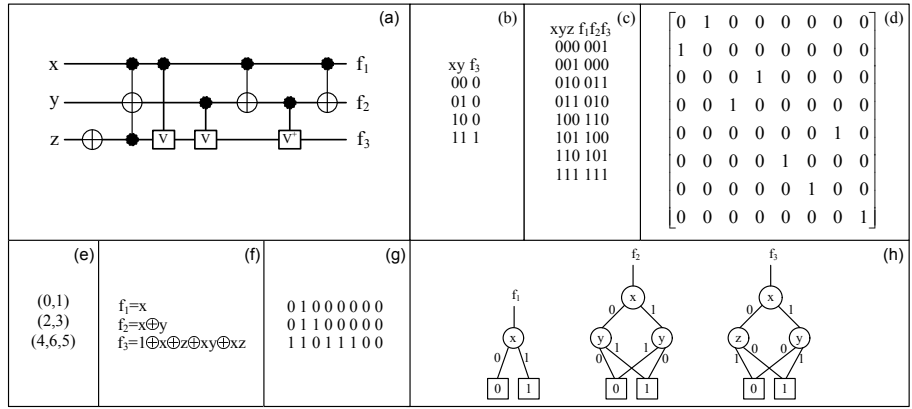}
\caption{A sample quantum circuit that implements a reversible specification (a), the specification in various formats: irreversible truth table (b), reversible truth table (c), matrix representation (d), cycle form (e), PPRM (f), RM spectrum (g), ROBDD (h).}
\label{fig:models}
\end{figure}

\vspace{1mm}
\textbf{Representation models}. Reversible functions can be described in several ways, as illustrated in Fig. \ref{fig:models}.
\begin{itemize}
    \item [$\bullet$] \emph{Truth tables}. The simplest method to describe a reversible function of size $n$ is a truth table with $n$ columns and $2^n$ rows.
    \item  [$\bullet$] \emph{Matrix representations}. A Boolean reversible function (permutation) $f$ can be described by a 0-1 matrix with a single 1 in each column and in each row (a \emph{permutation matrix}), where the non-zero element in row $i$ appears in column $f(i)$. A different matrix representation for linear functions \cite{PatelQIC08} is described in Section \ref{sec:asymp}.
    \item [$\bullet$] \emph{Reed-Muller expansion}. To denote a specification with algebraic formula, \emph{Positive polarity Reed-Muller} ({PPRM}) expansion can be applied. PPRM expansion uses only un-complemented variables and can be derived from the EXOR-Sum-of-Products ({ESOP}) description by replacing $a'$ with $a\oplus1$ for a complemented variable $a$. The PPRM expansion of a function is canonical\footnote{A canonical form is a way to rule out multiple representations of the same object. Given two different  representations, they can be converted to canonical forms. The objects are equivalent if and only if the canonical forms match.} and is defined as follows.
    \begin{equation}
    \begin{array}{l}
     f(x_1 ,x_2 ,...,x_n ) = a_0  \oplus a_1 x_1  \oplus  \cdots  \oplus a_n x_n  \oplus a_{12} x_1 x_2  \oplus  \cdots  \\
     \,\,\,\,\,\,\,\,\,\,\,\,\,\,\,\,\,\,\,\,\,\,\,\,\,\,\,\,\,\,\,\,\,\,\, \oplus a_{n,n - 1} x_{n - 1} x_n  \oplus  \ldots  \oplus a_{12...n} x_1 x_2  \cdots x_n  \\
     \end{array}
    \label{eq:pprm}
    \end{equation}
    A compact way to represent PPRM expansions is the vector of coefficients $a_0$, $a_1$, ..., $a_{12...n}$, called the \emph{RM spectrum} of the function. Consider an $n$-variable function and record its values (from the truth table) in a $2^n$-element bitvector $F$. Then, the RM spectrum ($R$) of $F$ over the two-element field\footnote{The finite field GF(2) consists of elements 0 and 1 for which addition and multiplication are equivalent to logical XOR and AND operations, respectively.} GF(2) is defined as $R=M^nF$ where
    \begin{equation}
    M^0=[1],\,\
    M^n=\left[ {\begin{array}{*{20}c}
       {M^{n - 1} } & 0  \\
       {M^{n - 1} } & {M^{n - 1} }  \\
    \end{array}} \right]
    \end{equation}
    \item [$\bullet$] \emph{Cycle expansion}. Viewing a reversible function as a permutation, one can represent it as a product of disjoint cycles.
    \item [$\bullet$] \emph{Decision Diagrams}. A reversible function can be represented by a \emph{Binary Decision Diagram} (BDD) \cite{Bryant86,Hachtel:2000}. A BDD is a directed acyclic graph where the Shannon decomposition (i.e., $f = x_i f_{x_i=0} + x_i f_{x_i=1}$) is applied on each non-terminal node. \citeN{Bryant86} proposed Reduced Ordered BDDs (ROBDDs), which offer canonical representations of Boolean functions. An ROBDD can be constructed from a BDD by ordering variables, merging equivalent sub-graphs and removing nodes with identical children. Several more specialized BDD variants have emerged for reversible and quantum circuits \cite{Viamontes09}. In general, a BDD of a function may need an exponential number of nodes. However, BDD variants can represent many practical functions with only polynomial numbers of nodes.
\end{itemize}

\textbf{Factorizations}. A given cycle of length $>2$ can be \emph{factorized} into smaller cycles. For example, the 4-cycle $(0,1,5,4)$ can be factorized into three 2-cycles $(0,1)(0,5)(0,4)$. A factorization is of \textit{type} $\alpha=(\alpha_2, \dots, \alpha_k)$ if it results in exactly $\alpha_2$ 2-cycles, $\alpha_3$ 3-cycles and so on. Define $\langle \alpha \rangle=\Sigma_{j\geq 2} (j-1)  \times \alpha_j$ for an $n$-bit permutation where $\alpha$ satisfies $\langle \alpha \rangle \geq n-1$. A factorization is \emph{minimal} if $\langle \alpha \rangle = n-1$. For instance, the factorization $(0,1,5)(0,4)(2,3)$ of $g=(0,1,5,4) (2,3)$ is of type $\alpha=(2,1)$ and is not minimal, $2 \times (2-1)+1 \times (3-1)=4>2$. Two factorizations are \emph{equivalent} if one can be obtained from the other by repeatedly exchanging adjacent factors that are disjoint. For example, factorizations $(0,1,5)(0,4)(2,3)$ and $(3,2)(5,0,1)(0,4)$ are equivalent. Since cycles $(0,1,5)$ and $(0,4)$ share a common element, they do not commute.
\subsection{Quantum Gates and Circuits}
A quantum bit, \emph{qubit}, can be treated as a mathematical object that represents a quantum state with two basic states $|0\rangle$ and $|1\rangle$. It can also carry a linear combination $|\psi\rangle = \alpha|0\rangle+\beta|1\rangle$ of its basic states, called a \emph{superposition}, where $\alpha$ and $\beta$ are complex numbers and $|\alpha|^2$+$|\beta|^2$=1. Although a qubit can carry any norm-preserving linear combination of its basic states, when a qubit is \emph{measured}, its state collapses into either $|0\rangle$ or $|1\rangle$ with probabilities $|\alpha|^2$ and $|\beta|^2$, respectively. A \emph{quantum register} of size $n$ is an ordered collection of $n$ qubits. Apart from the measurements that are commonly delayed until the end of a quantum computation, all quantum computations are reversible.

\begin{figure}[tb]
\centering
\includegraphics[height=22mm]{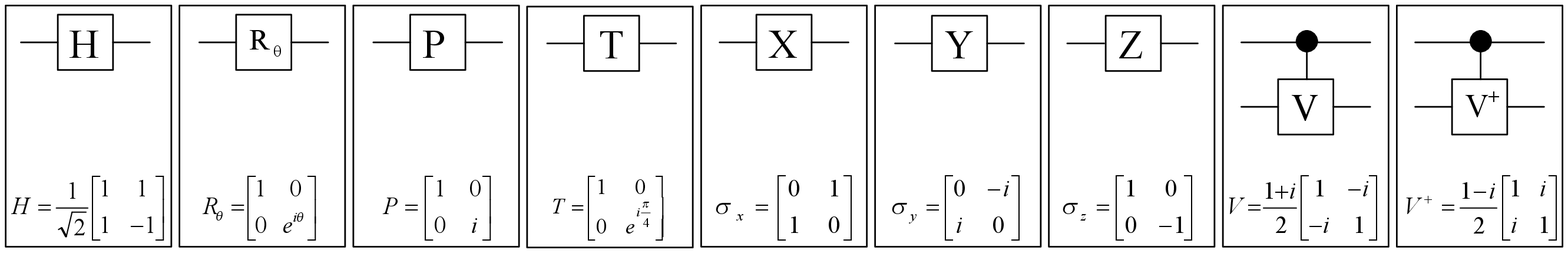}
	\caption {Basic quantum gates: Hadamard, phase shift, Phase, $\frac{\pi}{8}$, Pauli-X (NOT), Pauli-Y, Pauli-Z, controlled-V, and controlled-V$^\dag$.} \label{fig:Ggate}
\end{figure}

\vspace{1mm}
\textbf{Quantum gates}. A matrix $U$ is \emph{unitary} if $UU^\dag=I$ where $U^\dag$ is the conjugate transpose of $U$ and $I$ is the identity matrix. An $n$-qubit \emph{quantum gate} is a device which performs a $2^n\times2^n$ unitary operation $U$ on $n$ qubits in a specific period of time. For a gate $g$ with a unitary matrix $U_g$, its inverse gate $g^{-1}$ implements the unitary matrix $U_g^{-1}$. Among various quantum gates with different functionalities \cite{Nielsen00} are Hadamard (H), phase shift ($R_\theta$), controlled-V, controlled-V$^\dag$, and the Pauli gates, which are defined in Fig. \ref{fig:Ggate}. For $\theta=\pi/2$ ($e^{i\theta}=i$) and $\theta=\pi/4$, the phase shift gate is named the Phase (P) and $\frac{\pi}{8}$ (T) gates, respectively. NCV is the library of NOT, CNOT, controlled-V and controlled-V$^\dag$. For an arbitrary single-qubit gate $U$, a controlled-$U$ gate is a 2-qubit gate with one control and one target which applies $U$ on the target qubit whenever the control condition is satisfied. Basic quantum gates are illustrated in Fig. \ref{fig:Ggate}. The set of reversible gates is a subset of all possible quantum gates, distinguished by having only 0s and 1s as matrix elements. It would be misleading to call reversible circuits \emph{quantum} just because they are used in quantum information processing. As we show in Section \ref{sec:basic}, reversible circuits can be described and manipulated without leaving the Boolean domain. The size of reversible circuits can sometimes be reduced by introducing non-Boolean gates (Section \ref{sec:post_syn}).

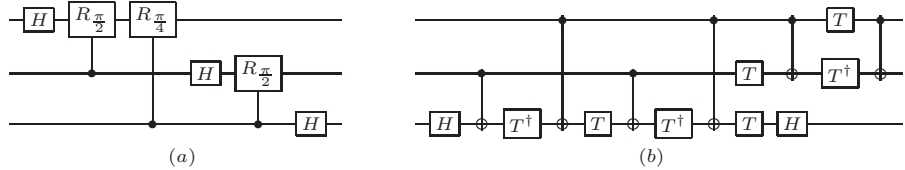
\begin{figure}
\scriptsize
\[
          \Qcircuit @R=1em @C=0.8em {
            &\gate{H} &\gate{R_{\frac{\pi}{2}}}	&\gate{R_{\frac{\pi}{4}}}&\qw&\qw&\qw&\qw
            & & & & && \qw & \qw & \qw
            & \control \qw & \qw &
            \qw & \qw & \control \qw & \qw & \control \qw & \gate{T} &
            \control \qw & \qw \\
            &\qw&\ctrl{-1}	&\qw&\gate{H}	&\gate{R_{\frac{\pi}{2}}}&\qw&\qw
            & && & & & \qw & \control \qw & \qw &
            \qw \qwx & \qw & \control \qw & \qw & \qw \qwx &
            \gate{T} & \targ \qwx & \gate{T^\dagger} & \targ \qwx  & \qw \\
             &\qw		&\qw&\ctrl{-2}&\qw&\ctrl{-1}&\gate{H}	&\qw
             &&  & & && \gate{H} & \targ \qwx &
            \gate{T^\dagger} & \targ \qwx & \gate{T} & \targ \qwx &
            \gate{T^\dagger} & \targ \qwx & \gate{T} & \gate{H} & \qw & \qw &
            \qw \\
            &&&&\lstick{(a)}&&&&&&&&&&&&&&&\lstick{(b)}&&&&&&
            }
\]
\centering
\caption{A three-qubit Quantum Fourier Transform (a), decomposing the Toffoli gate into one-qubit and six CNOT gates; six CNOT gates are required [Shende and Markov 2009] (b).}
\label{fig:QFTDCM}
\end{figure}

\vspace{1mm}
\textbf{Quantum circuits}. A \emph{quantum circuit} consists of quantum gates, interconnected by qubit carriers (i.e., wires) without feedback and explicit fanouts. Fig. \ref{fig:QFTDCM}a illustrates a 3-qubit quantum circuit for the Quantum Fourier Transform (QFT) which includes the Hadamard and phase shift gates. The inverse of a quantum circuit is constructed by inverting each gate and reversing their order. A set of gates is universal for quantum computation if any unitary operation can be approximated with arbitrary accuracy by a quantum circuit which contains only those gates. The gate library consisting of CNOT and single-qubit gates is universal for quantum computation \cite{Nielsen00}. Fig. \ref{fig:QFTDCM}b shows a decomposition of the Toffoli gate into H, T, T$^\dag$, and CNOT gates; six CNOTs are required for Toffoli \cite{Shende09}. The search space for quantum-logic synthesis is not finite, and circuits implementing generic unitary matrices require $\Omega(4^n)$ gates \cite{shende-2004}.

\vspace{1mm}
\textbf{Stabilizer circuits}. The gates Hadamard, Phase, and CNOT are called stabilizer gates. A stabilizer circuit is a quantum circuit consisting of stabilizer gates and measurement operations. Stabilizer circuits have applications in quantum error correction, quantum dense coding, and quantum teleportation \cite{Nielsen00}. According to the Gottesman-Knill theorem \cite{Nielsen00}, quantum circuits exclusively consisting of the following components can be efficiently simulated on a classical computer in polynomial time:
\begin{itemize}
    \item [$\bullet$] A state preparation N-circuit with initial value $|000...0\rangle$ --- qubit preparation in the computational basis,
    \item [$\bullet$] Quantum gates from the Clifford group (Hadamard, Phase, CNOT, and Pauli gates),
    \item [$\bullet$] Measurements in the computational basis
\end{itemize}

\textbf{Evaluation and simulation of quantum circuits}. For matrices $A_{m \times n}$ and $B_{p \times q}$, the tensor (Kronecker) product $A\otimes B$ is a matrix of size $mp \times nq$ in which each element of $A$ is replaced by $B$. The unitary matrix effected by several gates acting on disjoint qubits (in parallel) can be calculated as the tensor (Kronecker) product of gate matrices. For a set of $k$ gates $g_1$, $g_2$, ..., $g_k$ with matrices $U_1$, $U_2$, ..., $U_k$ cascaded in a quantum circuit $C$ (sequentially), the matrix of $C$ can be calculated as $U_k U_{k-1} ... U_1$. Straightforward simulation of quantum circuits by matrix multiplication requires $\Omega(2^n)$ time and space \cite{Viamontes09}. To improve runtime and memory usage, algorithmic techniques have been developed for high-performance simulation of quantum circuits \cite{Shi06,Viamontes09}.\footnote{\PP\,(Probabilistic Polynomial-Time) is the class of decision problems solvable by an \NP\, (Nondeterministic Polynomial-Time) machine which gives the correct answer (i.e., `Yes' or `No') with probability $>\frac{1}{2}$. \P$^\PP$\, (\P\, with \PP\, oracle) includes decision problems solvable in polynomial time with the help of an oracle for solving problems from \PP. Quantum circuit simulation belongs to the complexity class \P$^\PP$.}

\vspace{1mm}
\textbf{Quantum circuit technologies}. To physically implement qubits, different quantum-mechanical systems have been proposed, each with particular strengths and weaknesses, as discussed in the \emph{Quantum Computation Roadmap} (http://qist.lanl.gov/qcomp\_map.shtml). Leading candidate technologies represent the state of a qubit using
\begin{itemize}
  \item [$\bullet$] A two-level motion mode of a trapped ion or atom,
  \item [$\bullet$] Nuclear spin polarizations in nuclear magnetic resonance (NMR),
  \item [$\bullet$] Single electrons contained in Gallium arsenide (GaAs) quantum dots,
\item [$\bullet$] The excitation states of Josephson junctions in superconducting circuits,
\item [$\bullet$] The horizontal and vertical polarization states of single photons.
\end{itemize}
Quantum gates are effected by shining laser pulses on neighboring ions or atoms, applying electromagnetic pulses to spins in a strong magnetic field, changing voltages and/or current in a superconducting circuit, or passing photons through optical media. These and other technologies are discussed in textbooks \cite{Nielsen00} and research publications \cite{Politi09,Gao2010}.

\vspace{1mm}
\textbf{Interpreting quantum circuit diagrams}. Representing quantum circuits with circuit diagrams invites analogies with conventional CMOS circuits, but there are several fundamental differences.
\begin{enumerate}
    \item In a quantum circuit, qubits typically exist as fixed physical entities (e.g., electrons, photons or nuclei), and quantum gates operate on a qubit register (some gates can be invoked in parallel). This is in contrast to conventional semiconductor circuits where signals travel through gates, often fanning out and reconverging.
    \item Wirelines in a quantum circuit are used to trace the different states of a qubit during computation. Unlike in conventional circuits, wirelines in quantum and reversible circuits have sequential semantics. This can be illustrated by considering constant-propagation, i.e., simplifying a circuit when some of the inputs are given known values. Even when the input values are 0 or 1, wirelines in circuits like the reversible adder from \cite{Cuccaro:quant-ph0410184} (Fig. \ref{fig:copy}d) cannot be removed because they are also used to store intermediate values of computation.
    \item In many implementations, quantum gates are invoked by electromagnetic pulses, in which case the different gates of a combinational circuit appear for short periods of time and then disappear. This is in contrast to more familiar circuits in semiconductor chips, where independently existing gates are connected by metallic interconnect. Photonic quantum circuits use explicit interconnect in the form of photonic waveguides.
    \item Conventional circuits are typically synchronized through sequential elements (latches and flip-flops) because the timing of individual gates cannot be controlled accurately. In quantum circuits where each gate can be invoked at a precisely specified moment in time, there is no need for synchronization using sequential gates, and the entire computation can be scheduled by timing each combinational gate.
    \item In conventional circuits, each wire is assumed to carry a 0 or 1 signal, and each output of a combinational circuit is deterministically observable at the end of a clock cycle. However, these assumptions break down in a quantum circuit that generates non-Boolean values \cite{Nielsen00} because $(i)$ multiple qubits can be entangled, $(ii)$ to directly observe a qubit, it must be measured, which generates a nondeterministic outcome and affects other entangled qubits.
\end{enumerate}
These differences between quantum and conventional circuits are sometimes misunderstood in the literature, as we point out in Section \ref{sec:open}.
\subsection{Circuit Cost Models} \label{sec:cost_models}
Current quantum technologies suffer from intrinsic limitations which prohibit some circuits and favor others, prime examples are the small number of available qubits and the requirement that gates act only on geometrically adjacent qubits (in a particular layout). To be relevant in practice, circuit synthesis algorithms must be able to satisfy technology-specific constraints and improve technology-specific cost metrics. For example, currently popular trapped-ions \cite{TrappedIon} and liquid-state NMR \cite{Negrevergne06} technologies allow computation on sets of 8-12 qubits in a linear nearest neighbor (LNN) architecture where only adjacent qubits can interact. Furthermore, a physical qubit can hold its state only for a limited time, called \emph{coherence time}, which varies among different technologies from a few nanoseconds to several seconds \cite{Meter06}. Because of decoherence, qubits are fragile and may spontaneously change their joint states.

Just as in conventional circuits, the trivial \emph{gate count} metric does not adequately reflect the resources required by different gates. Similar to transistor counts, used to compare logic gates implemented in CMOS chips, one can define the technology-specific cost of quantum gates by decomposing them into elementary blocks supported by a particular technology. A physical implementation of an elementary operation depends on the Hamiltonian\footnote{A Hamiltonian describes time-dependent behavior of a quantum system and can be compared to a set of forces acting on a non-quantum system.}
of a given quantum system \cite{Zhang03}. For example in a one-dimensional exchange, i.e., Ising Hamiltonian characterized by interaction in the $z$ direction only, the 2-qubit SWAP gate requires three qubit interactions. In a two-dimensional exchange with the XY Hamiltonian, it can be implemented by a single two-qubit interaction. In an ion trap system, ``elementary gates'' are implemented with carefully tuned RF pulse sequences. Gate costs can be affected not only by direct resource requirements (size, runtime, available frequency channels) but also by considerations of circuit reliability in the context of frequent transient errors (e.g., decoherence of quantum bits). Some gates may be more amenable to error-correction than others, e.g., the CNOT gate and other linear transformations allow for convenient fault-tolerant extensions.
In order to abstract away specific technology details, several abstract cost functions have been proposed in the literature. However, their relevance strongly depends on future developments in quantum-circuit technologies.
\begin{itemize}
    \item [$\bullet$] \emph{Speed} was defined in \cite{Beckman96} to approximate the runtime of a quantum computation on an ion trap-based quantum technology, assuming all laser pulses take equal amounts of time. They observed that the C$^k$NOT gate ($k = 1, 2,$ etc.) can be implemented by $2k + 3$ laser pulses. The authors assumed that only one gate can be applied at a time.
    \item [$\bullet$] \emph{Number of one-qubit gates and CNOT} (or any other two-qubit gate) is a complexity metric for quantum synthesis algorithms. Since CNOT is a linear gate, the number of one-qubit gates (excluding inverters) needed to express a computation is defined as a measure of non-linearity for a given computation \cite{Shende09}.
    \item [$\bullet$] \emph{Quantum cost} (QC) is defined as the number of NOT, CNOT, controlled-V and controlled-V$^\dag$ gates required for implementing a given reversible function. These gates can be efficiently implemented in an NMR-based quantum technology by a sequence of electromagnetic pulses \cite{Lee06}. Under any other quantum technology, primitive gates can be adapted similarly. For example, while Toffoli needs five gates from the NCV library (two CNOT, two controlled-V, and one controlled-V$^\dag$ gates) it needs exactly six CNOTs and several one-qubit gates under the universal set of one-qubit and CNOT gates (Fig. \ref{fig:QFTDCM}b) \cite{Shende09}. In another example, the Fredkin gate is easier to implement than the Toffoli gate under some quantum technologies \cite{Fei02}.\footnote{Fredkin can be constructed using three Toffoli gates by adding one control to each CNOT gate in Fig. \ref{fig:Fig1}b. The three Toffolis can then be simplified into two CNOTs around a Toffoli.} A single-number cost model, based on the number of two-qubit operations required to simulate a given gate, was used in \cite{MaslovTCAD11} where costs of both $n$-qubit Toffoli and $n$-qubit Fredkin gates (and $n \geq 3$) are estimated as $10n-25$. QC of a circuit is calculated by a summation over the QCs of its gates.
    \item [$\bullet$] \emph{Interaction cost} is the distance between gate qubits for any 2-qubit gate. Quantum circuit technologies with 1D, 2D and 3D interactions exist \cite{Cheung07}. Interaction cost for a circuit is calculated by a summation over the interaction costs of its gates.
    \item [$\bullet$] \emph{Number of ancillae} and \emph{garbage bits} (ancillae not reset to 0) reflects the limited number of qubits available in contemporary quantum computers.
    \item [$\bullet$] \emph{Depth} (or the number of levels) is defined as the largest number of elementary gates on any path from inputs to outputs in a circuit. When any subset of gates can be invoked simultaneously, decreasing circuit depth reduces circuit latency. This assumption is trivial for conventional semiconductor circuits because the gates are manufactured individually and exist at the same time. However, when quantum gates are invoked by electromagnetic pulses, their parallel invocation must clear a number of obstacles --- it should be possible to select just the right set of qubits on which the gates are applied, which may require several laser sources and possibly several pre-determined wavelengths. When the parallel gates perform different functions, interference between them may limit achievable parallelism. Practical quantum computers can either apply the same gate to all qubits \emph{or} apply different gates to a small number of qubits.

\end{itemize}

As pointed out in \cite{Beckman96}, specific quantum-circuit technologies may entail more involved cost functions where the delay of a gate may depend on neighboring gates. The abstract cost functions introduced above do not capture such effects.
\begin{figure}[t]
\centering
\includegraphics[height=20mm]{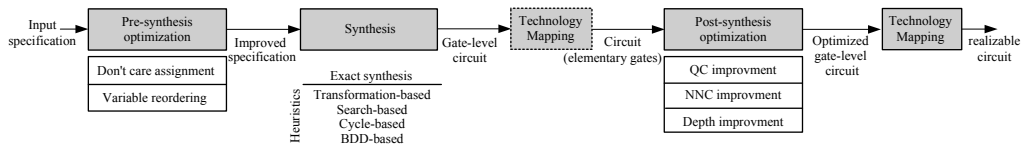}
\caption{A general flow used in recent reversible logic synthesis methods.}
\label{fig:flow}
\end{figure}

\section{Generation and Optimization of Reversible Circuits}  \label {sec:picture}
In this section we outline key steps in generation and optimization of reversible circuits, as illustrated in Fig. \ref{fig:flow}. Algorithmic details will be given in Sections \ref{sec:syn} and \ref{sec:post_syn}. To implement an irreversible specification using reversible gates, ancillae should be added to the original specification where the number of added lines, their values, and the ordering of output lines affect the cost of synthesized circuits. This process can be either performed prior to synthesis or in a unified approach during synthesis.\\

\textbf{Synthesis} seeks reversible circuits that satisfy a reversible specification. It can be performed optimally or heuristically.
\begin{itemize}
\item [$\bullet$] \emph{Optimal} iterated deepening A*-search (IDA*) algorithm was used in \cite{ShendeTCAD03} to find optimal circuits of all 3-input reversible functions. \citeN{GolubitskyDAC10} observed that an optimal realization of some reversible functions can be constructed from an optimal circuit of another function --- no need to synthesize all functions independently. For example, optimal circuits for $f^{-1}$ can be constructed by reversing optimal circuits for $f$. By exploiting such symmetries and using a hashing technique, the authors found optimal circuits for all 4-input permutations. Symbolic reachability analysis \cite{HungTCAD06} and Boolean satisfiability (SAT) \cite{GrosseTCAD09} have been applied to find optimal realizations for reversible functions. These methods mainly formulate the synthesis problem as a sequence of instances of standard decision problems, such as Boolean satisfiability, and use third-party software to solve these problem instances. Only a small number of qubits and gates can be handled by these methods.
\item [$\bullet$] \emph{Asymptotically optimal} synthesis was proposed by \citeN{PatelQIC08} for linear reversible circuits which leads to $\Theta(n^2/\log n)$ CNOT gates in the worst case. \citeN{Maslov07} addressed depth-optimal synthesis of \emph{stabilizer circuits} and proposed a synthesis algorithm that constructs circuits by concatenating $90n+O(1)$ stages, each stage containing only one type of gates (CNOTs or certain one-qubit gates). Asymptotically optimal methods may not produce optimal circuits for specific inputs.
\end{itemize}

Since most circuits of practical interest are non-linear and too large for optimal synthesis, heuristic algorithms were proposed. The choice of a representation model for reversible functions plays a significant role in developing effective synthesis algorithms. Each model favors certain types of reversible functions by representing them concisely. Synthesis algorithms are developed by detecting such simple cases and decomposing reversible functions into sequences of simpler functions in a given model.

\begin{itemize}
\item [$\bullet$] \emph{Transformation-based} methods \cite{MillerDAC03,MaslovTODAES07} iteratively select a gate so as to make a function's \emph{truth table or RM spectrum} more similar to the identity function. These methods are mainly efficient for permutations where output codewords follow a regular (repeating) pattern.
\item [$\bullet$] \emph{Search-based} methods \cite{GuptaTCAD06,DonaldJETC08} traverse a search tree to find a reasonably good circuit. These methods mainly use the \emph{PPRM expansion} to represent a reversible function. The efficiency of these methods is highly dependent on the number of circuit lines and the number of gates in the final circuit.
\item [$\bullet$] \emph{Cycle-based} methods \cite{ShendeTCAD03,SaeediJETC10} decompose a given permutation into a set of disjoint (often small) \emph{cycles} and synthesize individual cycles separately. Compared to other algorithms, these methods are mainly efficient for permutations without regular patterns and reversible functions that leave many input combinations unchanged.
\item [$\bullet$] \emph{BDD-based} methods \cite{WilleDAC09,WilleDAC10} use \emph{binary decision diagrams} to improve sharing between controls of reversible gates. These techniques scale better than others. However, they require a large number of ancilla qubits --- a valuable resource in fledgling quantum computers.
\end{itemize}

Several other heuristics do not directly use the discussed representation models. Some reuse algorithms developed for conventional logic synthesis, e.g., the algorithm proposed in \cite{ws:mp} uses ancillae to convert an optimized irreversible circuit into a reversible circuit. In \cite{FTR:2007} a circuit is constructed as a cascade of ESOP gates in the presence of some ancillae. Another approach uses abstract group theory to synthesize reversible circuits \cite{Storme:jucs_5_5,Rentergem:2007,YangLNCS06}.
However, as of 2011, empirical performance of reported implementations lags behind that of more established approaches. Heuristic synthesis is discussed in Section \ref{sec:Heuristics}, while synthesis of optimal circuits is explored in Sections \ref{sec:exact} and \ref{sec:asymp}.

\vspace{1mm}
\textbf{Post-synthesis optimization}. The results obtained by heuristic synthesis methods are often sub-optimal. Further improvements can be achieved by local optimization.

\begin{itemize}
\item [$\bullet$] \emph{Improving gate count and quantum cost}. To improve the quantum cost of a circuit, several techniques attempt to improve individual sub-circuits one at a time. Sub-circuit optimization may be performed based on offline synthesis of a set of functions using pre-computed tables \cite{PrasadJETC06,MaslovTCAD08}, online synthesis of candidates \cite{MaslovTODAES07,ArabzadehASPDAC10}, or circuit transformations that involve additional ancillae \cite{MillerISMVL10,MaslovTCAD11}.
\item [$\bullet$] \emph{Reducing circuit depth}. To realize a low-depth implementation of a given function, consecutive elementary gates with disjoint sets of control and target lines should be used to provide the possibility of parallel gate execution. Circuit depth may also be improved by restructuring controls and targets of different gates in a synthesized circuit \cite{MaslovTCAD08}.
\item [$\bullet$] \emph{Improving locality}. For the implementation of a given computation on a quantum architecture with restricted qubit interactions, one may use SWAP gates to move gate qubits towards each other as much as required. The interaction cost of a given computation can be hand-optimized for particular applications \cite{FDH:2004,Kutin07,Takahashi:2007}. A generic approach can also be used to either reduce the number of SWAP gates \cite{SaeediQIP11} or find the minimal number of SWAP gates \cite{HirataQIC11} for a circuit.
\end{itemize}

\begin{figure}[tb]
\centering
\includegraphics[height=40mm]{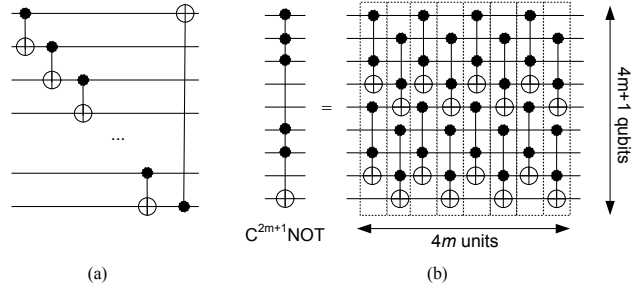}
	\caption {An optimal circuit used to illustrate limitations of local optimization [Prasad et al. 2006]. If the pattern is continued by adding one gate at a time until the circuit first becomes suboptimal, no local change could make the resulting circuit optimal (a), decomposition of a C$^{2m+1}$NOT gate [Asano and Ishii 2005], $m = 2^k$, $k\geq1$. In this figure, $k=1$ (b).} \label{fig:optlimit}
\end{figure}

Incremental optimization can significantly improve synthesis results, but it cannot guarantee optimality. To illustrate this, consider the NCT-optimal circuit in Fig. \ref{fig:optlimit}a \cite{PrasadJETC06}. Suppose the pattern is continued by adding one gate at a time until the circuit becomes suboptimal for the function it computes. In the resulting circuit, no suboptimal sub-circuits are formed, and hence no local-optimization method can find a reduction that is available. Section \ref{sec:post_syn} offers additional details on post-synthesis optimization.

\vspace{1mm}
\textbf{Technology mapping}. To physically implement a circuit using a given technology, all gates should be mapped (decomposed) into gates directly available in this technology. Such technology mapping can be applied either before post-synthesis optimization or after. \citeN{Barenco95} showed that a multiple-control Toffoli gate in a circuit on $n$ qubits can be mapped into a set of Toffoli gates, with different circuit sizes, depending on how many ancillae are available.

\begin{enumerate}
\item Without ancilla, $n \geq 3$: A C$^{n-1}$NOT gate can be simulated by $2^{n-1}-1$ controlled-V and controlled-V$^\dag$ gates and $2^{n-1}-2$ CNOTs.
\item With one ancilla, $n \geq 7$: A C$^{n-2}$NOT gate can be simulated by $8(n-5)$ Toffoli gates.
\item With $m-2$ ancillae, $m \in \{3,4, \cdots, \left\lceil n/2 \right\rceil\}$, $n \geq 5 $: A C$^m$NOT gate can be simulated by $4(m-2)$ Toffoli gates.
\end{enumerate}
\citeN{MaslovIEE03} converted Toffoli gates into (inverse) Peres gates which leads to $32m-96$ and $16m-32$ elementary gates from the NCV library for the cases (2) and (3), respectively.
\citeN{Asano05} presented a quantum circuit, illustrated in Fig. \ref{fig:optlimit}b, to simulate a C$^{2m+1}$NOT gate on $4m+1$ qubits, $m = 2^k$, $k\geq 1$, that contains $4m$ units of Toffoli gates. Each unit performs $m$ Toffoli operations simultaneously on $3m$ qubits. By eliminating individual-qubit manipulation, their circuit increases parallelism in quantum circuits at the cost of additional gates.

\begin{figure}[t]
\centering
\includegraphics[height=65mm]{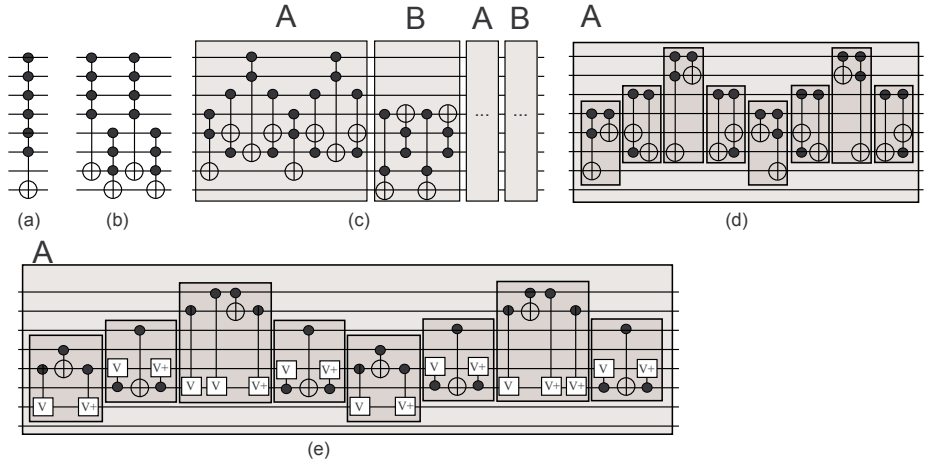}
\caption{Decomposition of a C$^6$NOT($a$,$b$,$c$,$d$,$e$,$f$,$h$) gate into smaller multiple-control Toffoli gates [Barenco et al. 1995] (b) and Toffoli gates [Barenco et al. 1995] (c), Peres gates [Maslov and Dueck 2003] (d), elementary gates [Maslov et al. 2008a] (e), in the presence of one garbage line.}
\label{fig:dcm}
\end{figure}

\citeN{MaslovTCAD08} improved the result of \cite{MaslovIEE03} by removing redundant controlled-V gates which leads to $12m-22$ and $24n-88$ gates for (2) and (3), correspondingly.
Fig. \ref{fig:dcm} illustrates the decomposition of a C$^6$NOT gate where b-c, d, and e are the results of applying the methods of \cite{Barenco95}, \cite{MaslovIEE03}, and \cite{MaslovTCAD08}, respectively. \citeN{Miller10} proposed techniques to reduce the number of elementary gates for C$^c$NOT, $c \in \{3,..., 15\}$ assuming $\{1, 2, ..., c-2\}$ ancillae.
Synthesis and post-synthesis optimization methods which consider the underlying gate libraries, e.g., to improve locality or to decrease circuit depth, should also benefit from an internal technology mapping.

\section{Algorithms for Reversible Circuit Synthesis}  \label {sec:syn}
In the following subsections, we discuss exact and asymptotically optimal synthesis methods followed by heuristic algorithms.

\subsection{Optimal Methods} \label {sec:exact}
For a reversible circuit with $n$ lines, where its optimal realization needs $h$ gates from a library $\mathcal{L}$, an enumerative method may branch $h$ ways on each $\mathcal{L}$-gate. For example, assume that only multiple-control Toffoli gates exist in the library. For this simplified case, an exhaustive method examines $(n \times 2 ^{n-1})^h$ gates\footnote{There are ${n \choose 1}$ possible NOT gates and ${n \choose 2}$ possible CNOT gates in which one of its two inputs can be the target output. Hence, the total number of 2$\times{n \choose 2}$ CNOT gates can be obtained. For a ($k$+1)-bit gate, $k \in (2, 3,\cdots, n-1)$, there are ${n-1 \choose k}$ possible gates when the target can be the $i$-th ($i \in [1,n]$) bit. Considering all possible bits as the target leads to $n \times {n-1 \choose k}$ ($k$+1)-bit gates. Therefore, the total number of gates is ${n \choose 1} + 2\times{n \choose 2} + n \times (\sum_{i \in (2\cdots n-1)} {n-1 \choose i})= n \times 2 ^{n-1}$.} to find an optimal circuit. For $n=3$, the worst-case circuit needs eight gates from the NCT library. Therefore, only $12^8$ different cases should be examined. For $n=4$, optimal circuits with 15 gates exist \cite{GolubitskyDAC10}, hence, $2^{75}\simeq3.8 \times 10^{22}$ different cases should be analyzed by exhaustive search to ensure that a min-cost circuit is found.

\vspace{1mm}
\textbf{3-qubit circuits}. \citeANP{ShendeTCAD03} performed gate-count optimal synthesis of 3-bit reversible functions by gradually building up a library of optimal circuits for all 8! permutations, rather than by dealing with each permutation individually \cite{ShendeTCAD03}. Noting that every sub-circuit of an optimal circuit is also optimal, they stored optimal circuits with $m$ gates and added one gate at the end of each stored circuit in all possible ways. Those resulting circuits that implement new functions can be added to the library. To lower memory usage when synthesizing a given permutation, instead of examining all optimal circuits with $k$ gates in the library for increasing values of $k$, the algorithm in \cite{ShendeTCAD03} stops at $m$ ($m \leq k$) gates and seeks circuits with $m+1$ gates that implement the permutation. In the absence of solutions, it seeks circuits with $m+2$ gates and so on.

\vspace{1mm}
\textbf{4-qubit circuits}. Optimal synthesis of 4-bit reversible functions was investigated in \cite{PrasadJETC06}, \cite{Yang:2008}, and \cite{GolubitskyDAC10}. Initially, \citeN{PrasadJETC06} introduced a data structure to represent all 40320 optimal 3-input and about 26,000,000 optimal 4-input reversible circuits with up to six gates from the NCT library. \citeN{Yang:2008} improved this method where the implementation of a specification on four variables was explored in a search tree based on a bidirectional approach \cite{MillerDAC03}.
Consequently, over 50\% of even 4-bit reversible circuits (approximately one quarter of all possible ones) were optimally realized with up to 12 NOT, CNOT and Peres gates. \citeN{GolubitskyDAC10} offered further improvements. They noted that in an optimal circuit with $k$ gates, the first $\lceil k/2 \rceil$ gates and the last $\lfloor k/2 \rfloor$ gates must also form optimal circuits for respective functions. Hence, they first synthesized all half-sized optimal circuits and stored them in a hash table.
The hash table was searched next for finding both halves of any optimal circuit with four inputs. Additionally, a simultaneous input/output relabeling (reordering) was applied, and symmetries of reversible functions were used to further reduce the search space. Optimal realization for the inverse $f^{-1}$ of a function $f$ was obtained by reversing an optimal circuit of $f$. The last two techniques reduce the search space by more than a factor of 48 (i.e., $2 \times 4!$). Running for less than 3 hours on a high-performance server with 16 AMD 2300 MHz processors and 64 GB RAM, \citeN{GolubitskyDAC10} found the distributions of gate-count optimal 4-bit circuits up to 15 gates reproduced in Table \ref{table:dist}.

\begin{table}[!t]
\tbl{The distribution of gate counts in gate-count optimal circuits for all 3- and 4-qubit functions with respect to the NCT library.\label{table:dist}}{
\centering
\begin{tabular}{|c|l|l|}
\hline
Number of & 3-bit Functions & 4-bit Functions \\
gates & \cite{ShendeTCAD03} & \cite{Golubitsky11}\\
\hline
\hline
15&0& 144\\
14&0& 37,481,795,636\\
13&0& 4,959,760,623,552\\
12&0& 10,690,104,057,901\\
11&0& 4,298,462,792,398\\
10&0& 819,182,578,179\\
9&0& 105,984,823,653\\
8&577& 10,804,681,959\\
7&10,253& 932,651,938\\
6&17,049& 70,763,560\\
5&8,921& 4,807,552\\
4&2,780& 294,507\\
3&625& 16,204\\
2&102& 784\\
1&12& 32\\
0&1 & 1 \\
\hline
Total & $2^3!$=40,320 &$2^4!=$20,922,789,888,000\\
\hline
\end{tabular}}
\end{table}

\vspace{1mm}
\textbf{Adapting algorithms from formal verification}. In order to find optimal circuits for reversible functions with more than four inputs, several sophisticated techniques draw upon algorithms and data structures from the field of formal verification \cite{Hachtel:2000}. Two optimal synthesis approaches for generic reversible and irreversible functions were developed in \cite{HungTCAD06} and \cite{GrosseTCAD09} where the former uses symbolic reachability analysis\footnote{Given a finite-state machine described by a sequential circuit and a set of states described by a property, the \emph{reachability problem} asks if the (un)desired states can be reached from the initial state through a sequence of valid transitions.} and the latter applies Boolean satisfiability. In \cite{HungTCAD06} a circuit is considered as a cascade of $L$ stages each of which is a 1-qubit or 2-qubit gate from the NCV library. Stage parameters (i.e., gate type and gate qubits) are modeled such that outputs of $i$-th stage are connected to inputs of $(i+1)$-th stage. In this scenario, a minimal-length circuit is equivalent to the smallest $L$. In contrast, for a given reversible function $f$, the algorithm of \cite{GrosseTCAD09} seeks the availability of a circuit implementing $f$ with a sequence of $d$ multiple-control Toffoli gates. Starting with $d=1$, $d$ is incremented until a circuit is found. While circuits are modeled in a similar fashion, the method in \cite{HungTCAD06} constructs an FSM (Finite State Machine) and employs a SAT solver to find a counter-example. To achieve this, instead of working with $L$ cascaded stages, $2^n$ parallel FSM instances are generated for truth table rows. The outputs of all $2^n$ instances at time $t$ are inputs of modules at time $t+1$. \citeN{GrosseTCAD09} used Boolean satisfiability and several common SAT techniques as well as problem-specific information to improve runtime. Optimal circuits with respect to interaction cost can be found similarly \cite{SaeediQIP11}. To improve runtime when handling large circuits, \citeN{WilleDATE08} used a generalization of Boolean satisfiability, Quantified Boolean Formula (QBF) satisfiability, and BDDs (Binary Decision Diagrams).
\subsection{Asymptotically Optimal Methods} \label{sec:asymp}
\citeN{Aaronson04} demonstrated that any stabilizer circuit can be restructured into 11 stages of Hadamard (H), Phase (P) and linear reversible circuits (C) in the order H-C-P-C-P-C-H-P-C-P-C. They also proved that the use of Hadamard and Phase provides at most a polynomial-time computational advantage since stabilizer circuits can be simulated by only NOT and CNOT gates. However, even when Hadamard and Phase gates are used, the size of a stabilizer circuit is likely to be dominated by the size of CNOT blocks. We therefore turn our attention to asymptotically optimal\footnote{An algorithm is asymptotically optimal if it performs at worst a constant factor worse than the best possible algorithm. Formally, for a problem which needs $\Omega(f(n))$ overhead according to a lower-bound theorem, an algorithm which requires $O(f(n))$ overhead is asymptotically optimal. While an such a method cannot find the solution optimally, no algorithm can outperform it by more than a fixed constant factor. On the other hand, other algorithms may find smaller circuits in specific cases, run faster or use less memory.} synthesis of linear functions.

When reversible functions are captured by (unitary) matrices, each row and each column include a single `1' and `0's elsewhere. A different model proposed in \cite{PatelQIC08} is specific to linear circuits and represents CNOT$(i,j)$ by inserting a `1' into the $(i,j)$ element of the identity matrix. This model allows one to cast synthesis of linear circuits as the task of reducing a given matrix (of the function to be synthesized) to the identity matrix by elementary row operations over GF(2). Each row operation corresponds to a CNOT gate, and the sequence of row operations gives a reversible circuit. This task is usually solved using Gaussian elimination, which requires $O(n^2)$ row operations and $O(n^3)$ time. To this end, the input matrix is reduced to an upper triangular matrix by a set of row operations, the resulting matrix is transposed, and this process is repeated on the transposed matrix. To reduce the number of gates, \citeN{PatelQIC08} partition an $n \times n$ matrix into a set of sections each one contains $m$ (e.g., $m=\log_2 n$) columns. To construct an upper-triangular matrix, the algorithm eliminates repeated rows in each section by applying carefully-planned row operations first. Then, diagonal entries are fixed, and Gaussian elimination is used to remove all off-diagonal entries. As in the standard approach, the same scenario is applied to the transposed matrix. This technique reduces the worst-case number of operations (equivalently the size of an $n$-wire CNOT circuit) to $\Theta(n^2/\log n)$ which is asymptotically optimal. Its runtime is improved to $O(n^3/\log n)$ versus $O(n^3)$ for Gaussian elimination.
\citeN{Maslov07} studied the depth (instead of size) of stabilizer circuits where only adjacent qubits can interact. By presenting a constructive algorithm based on Gauss-Jordan elimination, he demonstrated that any stabilizer circuit can be executed in at most $30n+O(1)$ stages composed of only generic two-qubit gates. For the library of CNOT and single-qubit gates, an (asymptotically-optimal) upper bound is $90n+O(1)$.
\subsection{Heuristic Methods} \label{sec:Heuristics}
Finding an optimal circuit for a given arbitrary-size reversible specification is intractable in general, hence heuristic methods have been developed to find reasonable solutions in practice. In this section, we review those methods that either significantly improved upon prior results or introduced new insights.

\begin{figure}[tb]
\centering
\includegraphics[height=55mm]{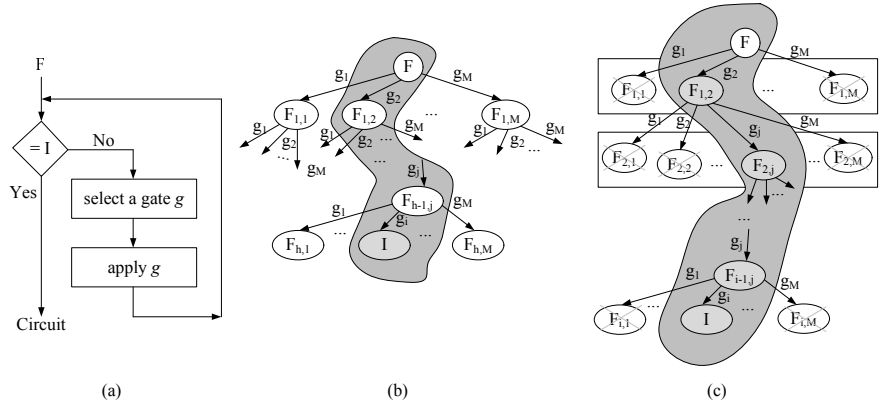}
\caption{Outlines of transformation-based algorithms (a), search-based methods (b), and search-based methods directed by a complexity metric (c). In this figure, $F$ is the input permutation, $I$ is the identity function, $g_i$ is a possible gate, $M$ is the maximum number of gates, and $F_{i,j}$ is a permutation which results from applying $g_j$ at the $i$-th level. }
\label{fig:search}
\end{figure}

\vspace{1mm}
\textbf{Transformation-based methods}. \citeN{MillerDAC03} proposed a synthesis method that compares the identity function ($I$) with a given permutation ($F$), as illustrated in Fig. \ref{fig:search}a, and applies reversible gates to \emph{transform} $F$ into $I$. To direct the transformation (or select a gate), the complexity metric used is the sum of Hamming distances between binary patterns of $F$ and $I$ at each truth table row. The algorithm iterates through the rows of the truth table, looks for differences between $F$ and $I$, and corrects these differences by applying multiple-control Toffoli gates  with positive controls only. This algorithm was improved in \cite{MaslovTODAES07} where the authors direct synthesis by the complexity of the Reed-Muller spectra
instead of the Hamming distance. The algorithm proposed in \cite{MaslovTODAES07} produces best-known circuits for several families of benchmark functions with regular patterns in their permutations.

Multiple-control Toffoli gates with both positive and negative controls in a column-wise (vs. row-wise as in \cite{MillerDAC03}) scenario were used in \cite{SaeediICCAD07}. This algorithm results in circuits composed of complex gates with common targets. Gates that share targets/controls can be further optimized by post-processing \cite{ArabzadehASPDAC10,MaslovTCAD11}.

\vspace{1mm}
\textbf{Search-based methods}. As shown in  Fig. \ref{fig:search}b the search process can be represented by a tree. One may \emph{search} for an implementation of a function by starting from an initial specification (root of the tree), applying individual gates (to generate branches), and repeating this process on the resulting functions until the identity specification is found in a branch. Given enough memory and time, this method can find a minimal circuit. It is useful when gate-counts and the numbers of inputs/outputs are small. To make this approach practical, one can select only those gates that minimize a specific metric as illustrated in Fig. \ref{fig:search}c. For example, in \cite{GuptaTCAD06} common sub-expressions between the PPRM expansions of multiple outputs are identified and used to simplify the outputs at each stage. Discovered factors are substituted into the PPRM expansions to determine their potential for leading to a solution where the primary objective is gate count (i.e., number of factors) minimization and the secondary objective is gate size (i.e., number of literals in each factor) reduction. To share factors among multiple outputs, candidate factors are selected among common sub-expressions in PPRM expansions. However, there is no guarantee that the resulting PPRM expressions contain fewer terms \cite{SaeediISVLSI07}. To relax optimization criteria, instead of evaluating previously substituted factors before new substitutions, \citeN{SaeediISVLSI07} considered all new factors first and proposed a hybrid method that applies the second approach before the first. \citeN{DonaldJETC08} improved the method of \cite{GuptaTCAD06} to handle gates in the NCTSFP library in a similar search-based framework. These algorithms can handle various gate types. Their performance is affected by the number of input bits and the size of resulting circuits.

\begin{figure}[tb]
\centering
\includegraphics[height=42mm]{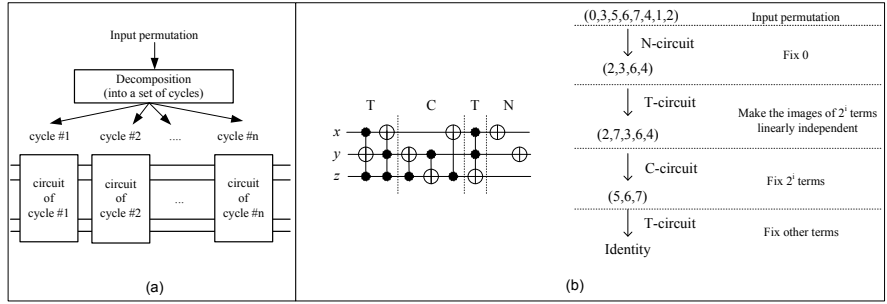}
\caption{A general outline of cycle-based methods (a), an example of the T$|$C$|$T$|$N method (b).}
\label{fig:Cycle}
\end{figure}

\vspace{1mm}
\textbf{Cycle-based methods}. Instead of working with an entire permutation, one can factor it into a set of \emph{cycles} and synthesize the resulting cycles separately as illustrated in Fig. \ref{fig:Cycle}a. This divide-and-conquer approach is particularly successful with reversible transformations that leave many inputs unchanged (sparse transformations).

\citeN{ShendeTCAD03} proposed an NCT-based synthesis method which applies NOT (N), Toffoli (T), CNOT (C), and Toffoli (T) gates in order (i.e., the T$|$C$|$T$|$N method) to synthesize a permutation. As illustrated in Fig. \ref{fig:Cycle}b, in the first C$|$T$|$N part, the terms $0$ and $2^i$ of a given function are positioned at their right locations. The last Toffoli circuit fixes the other truth table terms by decomposing the resulting permutation into a set of transpositions. Subsequently, each pair of disjoint transpositions is implemented by a synthesis algorithm separately, and the final circuit is constructed by cascading individual circuits. A similar method was introduced in \cite{YangLNCS06} except for working with neighboring 3-cycles, i.e., cycles whose elements differ only in two bits. This technique often produces an unnecessarily large number of cycles. An extension of method from \cite{ShendeTCAD03} described in \cite{PrasadJETC06} reduces synthesis cost by applying NOT and CNOT instead of Toffoli in many situations.

\citeN{SaeediJETC10} developed $k$-cycle synthesis, leading to significant reductions in the quantum cost for large cycles, based on seven building blocks --- a pair of 2-cycles, a single 3-cycle, a pair of 3-cycles, a single 5-cycle, a pair of 5-cycles, a single 2-cycle (4-cycle) followed by a single 4-cycle (2-cycle), and a pair of 4-cycles --- and a set of algorithms to synthesize a given cycle of length less than six \cite{SaeediJETC10}. Larger cycles are factorized into proposed building blocks. A hybrid synthesis framework was suggested which uses the cycle-based approach for irregular functions in conjunction with the method of \cite{MaslovTODAES07} for regular functions. The proposed cycle-based method leads to best-known circuits with respect to quantum cost for permutations which have no regular pattern. In addition, the maximum number of elementary gates for any permutation function in \cite{SaeediMEJ10} is less than $8.5n2^n + o(2^n)$, which is the sharpest upper bound for reversible functions so far (the lower bound is $n2^n/ \log n$ \cite{ShendeTCAD03}). A more efficient decomposition algorithm was proposed in \cite{SaeediMEJ10} which produces all minimal and inequivalent factorizations each of which contains the maximum number of disjoint cycles. These decompositions are used in a cycle-assignment algorithm based on the graph matching problem to select the best possible cycle pairs during synthesis.

\begin{figure}[tb]
\centering
\includegraphics[height=30mm]{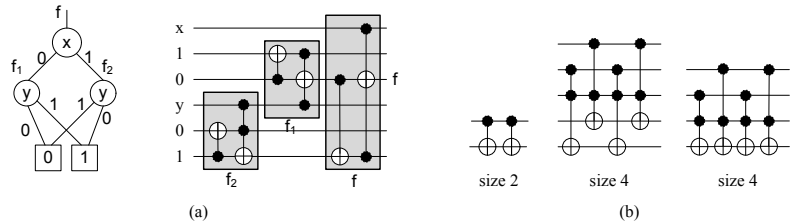}
\caption{A BDD-based circuit synthesis technique by example [Wille and Drechsler 2009] (a), templates of sizes 2 and 4 [Maslov et al. 2007] (b).}
\label{fig:BDD}
\end{figure}

\vspace{1mm}
\textbf{BDD-based methods}. \citeN{KerntopfDAC04} introduced a synthesis algorithm that uses binary decision diagrams (BDDs), and seeks to minimize the number of non-terminal DD nodes. At each step, all possible gates are examined, and the corresponding decision diagrams are constructed. The gates that minimize the complexity metric are selected and further analyzed by repeating the same process. \citeN{WilleDAC09} introduced a different algorithm that starts by constructing a BDD. Each BDD node is substituted by a cascade of reversible gates as shown in Fig. \ref{fig:BDD}a. Node sharing due to reduction rules in ROBDDs can cause gate fanout which is prohibited in reversible logic. To overcome this obstacle, the algorithm adds constant bits to emulate fanout --- in the worst case, for each BBD node, a new constant line may be added. While this algorithm leads to a good reduction in both quantum cost and runtime, many constant and garbage bits are added which makes the results impractical for quantum computers with a limited number of qubits. \citeN{WilleDAC10} reduced the number of lines by a post-processing technique where garbage lines are merged with appropriate constant lines. Although BDD-based techniques for reversible synthesis scale better than most other approaches, the large number of ancillae they generate makes the results difficult to use in practice, and the effort to consolidate ancillae can be substantial.
\section{Post-Synthesis Optimization}  \label {sec:post_syn}
To improve the results of synthesis algorithms, several optimization methods consider connected subsets of gates (sub-circuits) in a given circuit. Such sub-circuits are analyzed one by one and replaced by equivalent (smaller) sub-circuits to improve cost. This sub-circuit replacement approach can leverage earlier-discussed techniques
to improve large circuits using {\em peephole optimization} with linear runtime \cite{PrasadJETC06}.

\subsection {Quantum Cost Improvement} \label {sec:qc_imp}
Equivalent sub-circuits can be found using either windowing or sub-circuit optimization and replacement \cite{PrasadJETC06,MaslovTODAES07,MaslovTCAD11}.

\vspace{1mm}
\textbf{Library-based optimization}. \citeN{PrasadJETC06} proposed an algorithm that uses a large database of optimal circuits and seeks sub-circuits that can be replaced by smaller equivalent sub-circuits. In practice, the stored sub-circuits are likely to be very limited in size. \citeN{PrasadJETC06} introduced a compact data structure that can store all 3-bit reversible circuits and many 4-bit circuits with less than six gates. A windowing strategy proposed in \cite{PrasadJETC06} to identify contiguous sub-circuits can reorder some gates (without changing the overall functionality) to assemble larger 4-bit sub-circuits. The functionality of the sub-circuit found is computed, and a database look-up is performed to find an optimal circuit that implements the same functionality. The sub-circuit is replaced if this improves cost. Originally, this algorithm was applied to optimize reversible circuits   composed of NOT, CNOT and Toffoli gates, but it can work with other gates as well. Such optimizations rely heavily on a database of optimal implementations and an efficient windowing strategy.

Each circuit stored in a library can be viewed as a rule that simplifies any other circuit that computes the same function. For example, pairs of inverters, pairs of CNOTs and pairs of Toffoli gates cancel out because the same function can be computed by an empty circuit. To reduce the size of a library, such rules can be generalized by local circuit transformations, leading to more compact rule sets.

\vspace{1mm}
\textbf{Transformation rules and template-based optimization}. The work performed in \cite{IwamaDAC02} introduced the idea of local transformation of reversible circuits. While the main purpose of this work was not post-synthesis optimization, its results were extended by other researchers to improve circuit cost. The authors defined a canonical form for circuits in the NCT library, and introduced a complete set of rules to transform any NCT-constructible circuit into its canonical form, which may or may not be compact.

The concept of applying a rule set was extended in \cite{MillerDAC03} where the authors introduced several transformation rules based on a set of predefined patterns called \emph{templates}. A template $T$ is a reversible circuit that implements the identity function, which contains $m$ gates $g_1$, $g_2$, $\cdots$, $g_m$. For a circuit with multiple-control Toffoli and Fredkin gates, consider the first $k$ ($k > m/2$) gates of $T$ (i.e., $g_1$, $g_2$, $\cdots$ $g_k$). Suppose that these gates are found in a reversible circuit in sequence. It can be verified that the set of $m-k$ gates $g_m$, $\cdots$, $g_{k+2}$, $g_{k+1}$ can be applied instead of the initial $g_1$, $g_2$, $\cdots$ $g_k$ gates to reduce the gate count from $k$ to $m-k$.\footnote{If $g_1 \cdots g_k g_{k+1} \cdots g_m=I$, then $g_1 \cdots g_k=g_m^{-1} \cdots g_{k+1}^{-1}$. For self-inverse gates (Toffoli, Fredkin), $g^{-1}=g$.} The authors showed that applying the template matching method (called \emph{template application algorithm}) with two- and three-input templates only can improve the circuits.

In \cite{MaslovTCAD05}, template matching with up to six gates was used in post-synthesis optimization. The authors showed that there are 0, 1, 0, 1, 1, and 4 templates for gates with 1, 2, 3, 4, 5, and 6 gates, respectively. Their analysis shows that these seven templates comprise a complete set of templates of size $<6$, for $<4$ inputs. Similarly, the Toffoli-Fredkin templates were explored in \cite{MaslovTVLSI05} where the authors showed that there are 0, 1, 0, 3, 1, and 1 Toffoli-Fredkin templates for gates with 1, 2, 3, 4, 5, and 6 gates. Toffoli templates were extended in \cite{MaslovTODAES07} by the addition of all templates of size 7 (five templates) and a set of templates of size 9 (four templates). Fig. \ref{fig:BDD}b shows templates of sizes 2 and 4. In addition, the template application algorithm was enhanced leading to two templates of size 4 (vs. 1) and three templates of size 6 (vs. 4). \citeN{SaeediQIP11} extended the templates to work with up to three SWAP gates. Template-based optimizations can be time-consuming, but scale to large circuits due to their local nature \cite{MaslovTODAES07}. One can restrict template application to small subsets of gates and lines to improve runtime. Such post-processing can be used in peephole optimization with guaranteed linear runtime \cite{PrasadJETC06}.

\begin{figure}[tb]
\centering
\includegraphics[height=30mm]{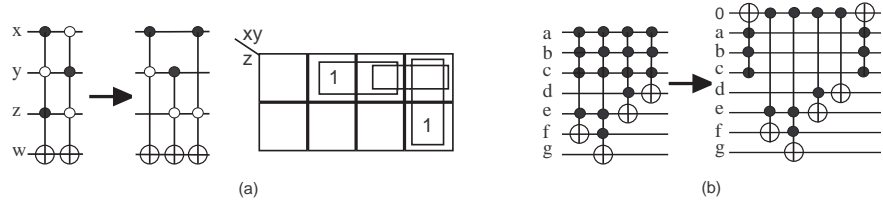}
\caption{K-map-based optimization [Arabzadeh et al. 2010] (a), ancilla insertion [Miller et al. 2010] (b).}
\label{fig:kmap}
\end{figure}

\citeN{ArabzadehASPDAC10} proposed a set of simplification rules in terms of positively and negatively controlled Toffoli gates. To optimize a sub-circuit which has gates with identical target as illustrated in Fig. \ref{fig:kmap}a, a C$^{n-1}$NOT gate is represented by a Boolean expression with $n-1$ inputs and one output where gate controls act as the inputs and the target behaves as the output. Next, this gate fills one cell of a Karnaugh map (K-map) of size $n$ (i.e., $n-1$ inputs, one output). To extract a simplified circuit, one can use a K-map cell clustering similar to the one used in irreversible logic. The authors showed that each cell with the value 1 can be used in an odd number of groups and each cell with the value 0 can be used in an even number of groups. Some templates in \cite{MaslovTCAD05}, e.g., the ones in Fig. \ref{fig:BDD}, can be regenerated by applying a set of simplification rules from \cite{ArabzadehASPDAC10}. This simplification approach is suitable for methods that generate subsequent gates on the same target line \cite{ws:mp,SaeediICCAD07}. An optimization in \cite{Soeken10} uses a window to select potential sub-circuits first. Then, re-synthesis, exact synthesis and template matching methods are applied to improve the selected sub-circuits.

\vspace{1mm}
\textbf{Qubit insertion}. Circuits can be simplified by adding ancillae. A well-known example is the implementation of the $n$-bit multiple-control Toffoli gate discussed in Section \ref{sec:picture}. Another example, the generic algorithm in \cite{MillerISMVL10} searches sub-circuits with a set of shared controls $C$. Their gates are simplified by removing controls in $C$. Two identical multiple-control Toffoli gates are inserted before and after the simplified sub-circuit as illustrated in Fig. \ref{fig:kmap}b where their controls are the qubits in $C$ and targets are on the zero-initialized line. This modification produces an equivalent but smaller circuit if the cost of added gates is smaller than that of removed controls in the multiple-control gates. This idea was further extended in \cite{MillerISMVL10} to add multiple ancillae.

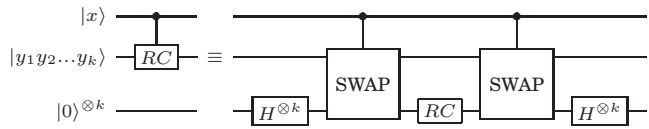
\begin{figure}
\scriptsize
\[
\Qcircuit @C=1em @R=1.5em {
\lstick{\ket{x}}& \ctrl{1}  &\qw& \;&&\qw& \ctrl{1} \qw  & \qw    & \ctrl{1}           		 & \qw & \qw  \\
\lstick{\ket{y_1y_2...y_k}}   & \gate{RC} &\qw& \equiv &&\qw& \multigate{1}{\text{SWAP}} 	& \qw      & \multigate{1}{\text{SWAP}} & \qw & \qw \\
\lstick{\ket{0}^{\otimes k}}  & \qw       &\qw& \;     &&\gate{H^{\otimes k}} & \ghost{\text{SWAP}}        	& \gate{RC}& \ghost{\text{SWAP}}        & \gate{{H}^{\otimes 
k}} & \qw \\
}
\]
\centering
\caption{Circuit equivalence used in [Maslov and Saeedi 2011].}
\label{fig:booleandomain}
\end{figure}

To compute a Boolean function by a quantum circuit, it is common to use only reversible (non-quantum) gates. However, the use of quantum gates offers more freedom and may facilitate smaller circuits in some cases. \citeN{MaslovTCAD11} proposed a circuit optimization that uses quantum Hadamard gates and therefore ventures beyond the Boolean domain. For a reversible circuit $RC$ and $\ket{00...0}$ ancillae, they consider the transformation $\ket{x}\ket{00...0} \mapsto RC\ket{x}\ket{00...0}$ with at most $n$ ancilla for $n$ primary inputs in the original reversible circuit. The ancillae are prepared by a layer of Hadamard gates, as shown in Fig. \ref{fig:booleandomain}. Sets of adjacent gates with shared     controls are identified. Since $H^{\otimes k}\ket{00...0}$ is a 1-eigenvector
of any $0$-$1$ unitary matrix $RC$, applying $RC$ to this eigenvector does not modify the state. After that, the shared controls are removed from the gates involved. The values are transferred to the ancillae by applying a set of Fredkin gates, and returned to the main qubits by reapplying the same set of Fredkin gates in the reverse order. This optimization is applied opportunistically wherever it improves circuit cost. It is particularly suitable for reversible circuits with many complex gates which can be easily reordered, such as those produced in \cite{ws:mp}.

\subsection{Reducing Circuit Depth}  \label{sec:depth}
Parallel circuits speed up computation and can tolerate smaller coherence times.\footnote{The study of parallel quantum algorithms has attracted attention in complexity theory too. \NC$_i$ is the class of decision problems solvable by a uniform family of Boolean circuits, with polynomial size, depth O($\log ^i (n)$), and fan-in 2. \QNC$^0$ is the class of constant-depth quantum circuits without fanout gates. The question whether \P\, $\subset$ \NC$_i$ or \P\, $\subset$ \QNC$^0$ is open.} \citeN{MaslovTCAD08} introduced a level compaction algorithm to reduce circuit level (or depth) of synthesized circuits by employing templates. To this end, a greedy algorithm was proposed which assigns an undefined level to all gates initially. Next, for each level $i$ the leftmost gate with an undefined level is examined to verify whether this gate can be executed at level $i$ or not. This process is continued until the algorithm finds no gate for execution at the $i$-th level. Next, a set of templates is applied, to change the control and target lines of different gates, and the level assignment process is repeated with the hope of improving circuit depth. Finally, $i$ is incremented and other gates are examined similarly. While the proposed method is useful for level compaction, its efficiency can be improved by applying a more efficient gate selection method.

\subsection{Improving Locality}\label{sec:Locality}
Quantum-circuit technologies often require that each gate involve only geometrically adjacent qubits (in a particular physical layout). Given a fixed number of qubits, a quantum architecture can be described by a simple connected graph $G = (V,E)$, where the vertices $V$ represent qubits and edges $E$ represent adjacent qubit pairs where gates can be applied \cite{Cheung07}. A complete graph, $K_n$, expresses the absence of constraints. The LNN (Linear Nearest Neighbor) architecture corresponds to a graph with $n$ vertices $v_1, \cdots, v_n$ where an edge exists between only neighboring vertices $v_i$ and $v_{i+1}$ for $1 \leq i < n$. Several systems of trapped ions \cite{TrappedIon}, liquid NMR (e.g.,~\cite{NMR}), and the original Kane model \cite{Kane} have been designed based on the interactions between linear nearest neighbor qubits. Two-dimensional square lattices (2DSL) corresponds to a graph on a two-dimensional Manhattan grid where only four neighboring qubits can interact. The relevant proposals for 2DSL include arrays of trapped ions \cite{TrappedIon}, Kane's architecture \cite{Skinner03}, and Josephson junctions \cite{PhysRevB.69.214501}. The three-dimensional square lattices (3DSL) model is a set of stacked 2D lattices where a qubit can interact with six neighboring qubits. 3DSL is less restrictive, but suffers from the difficulty of controlling 3D qubits. The architecture proposed in \cite{PhysRevLett.97.100501} relies on the 3DSL model. \citeN{Cheung07} introduced other architectures including the Star architecture with one vertex of degree $n- 1$ connected to all other vertices, and the Cycle ($C_n$), which is LNN with one extra interaction between the first and last qubit. The $k$-th power of the graph $G$, denoted by $G^k$ such as LNN$^k$, is the graph over the same vertex set (of $G$) with edges representing paths of length $k$ in $G$.

\vspace{1mm}
\textbf{SWAP insertion}. A naive method to satisfy (architectural) qubit-interaction constraints is to use SWAP gates in front of an improper gate $g$ to `move' the control (target) line of $g$ towards the target (control) line as much as required. Subsequently, SWAP gates should be added to restore the original ordering of circuit lines. This process can be repeated for all gates. More efficient circuits were found in application-specific studies that explored the physical implementations of the quantum Fourier transform \cite{Takahashi:2007,Maslov07}, Shor's factorization algorithm \cite{FDH:2004,Kutin:2007}, quantum addition \cite{Choi08}, and quantum error correction \cite{Fowler} for the LNN architectures. Researchers considered the impact of LNN constraints on the synthesis of general quantum/reversible circuits in \cite{Shende06} where their number of gates was increased by almost an order of magnitude, and in \cite{Mottonen06} and \cite{SaeediQIC11} where their numbers of CNOT gates were increased by at most a factor of $<2$. \citeN{Cheung07} discussed the translation overhead for converting an arbitrary circuit from one architecture to another one. Particularly, they showed that translating a circuit from $K_n$ to Star, LNN, 2DSL, and 3DSL requires $O(n)$, $O(n)$, $O(\sqrt{n})$, and $O(\sqrt[3]{n})$ overhead, respectively. Converting Star, 2DSL, and 3DSL, LNN$^{k}$, $C_n$, and $C^k_n$ to LNN requires $O(1)$, $O(\sqrt{n})$, $O(\sqrt[3]{n^2})$, $O(k)$, $O(1)$, and $O(1)$ overhead, respectively. Most importantly, Star is the weakest architecture among those considered, e.g., the overhead of converting a circuit from LNN to Star is $O(n)$.

\begin{figure}[t]
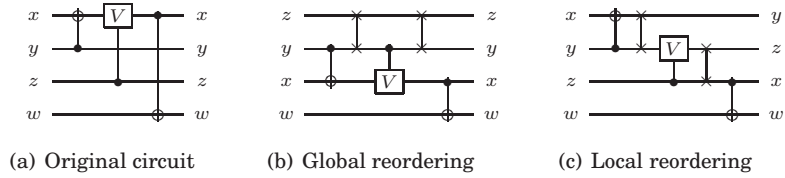

        \scriptsize
        \centering
        \subfigure[Original circuit\label{fig:reordering1}]{
          \input{figures/reordering_local1.tex}
        }\subfigure[Global reordering\label{fig:reordering_global2}]{
          \input{figures/reordering_global2.tex}
        }\subfigure[Local reordering\label{fig:reordering_local2}]{
          \input{figures/reordering_local2.tex}
        }
        \caption{Global and local reordering [Saeedi et al. 2011b].}%
        \label{Fig:reordering}
\end{figure}

\vspace{1mm}
\textbf{SWAP optimization}. To adapt circuits to restricted architectures, synthesis algorithms can minimize the number of elementary gates or the SWAP gates. In this context, exact and heuristic synthesis algorithms as well as post-synthesis optimization methods can be applied. Unlike optimal methods, heuristic post-synthesis optimizations scale well to large functions. Template matching for SWAP reduction and reordering strategies, \emph{global} and \emph{local} reordering, were introduced as powerful tools for SWAP reduction in \cite{SaeediQIP11}. In global reordering, lines with the highest interaction impact are sequentially chosen for reordering and placed at the middle line. This procedure is repeated until the cost cannot be reduced. In contrast, local reordering traverses a given circuit from inputs to outputs and adds SWAP gates only in front of each non-local gate, but not after. Instead, the resulting ordering is used for the rest of the circuit. This process is repeated until all gates are traversed, as illustrated in Fig. \ref{Fig:reordering}. Similar reordering scenarios were applied by hand to reduce the number of SWAPs in specific circuits, e.g., in \cite{Takahashi:2007} for QFT.

\vspace{1mm}
\textbf{Ensuring the minimal possible number of SWAP gates}. For a qubit set $\{x_1, x_2, \cdots, x_n\}$, assume that 1st, 2nd, ..., and $n$-th qubits should be placed at $C(x_1)$, $C(x_2)$,..., and $C(x_n)$ positions (C is the transformation function) to make a gate local. \citeN{HirataQIC11} showed the number of SWAP gates necessary for this purpose is at least the number of pairs in the set $S=\{(x_i, x_j)| C(x_j) <  C(x_i), i<j \}$ and a bubble sort generates this minimum number of SWAPs for each gate. To find the minimal number of SWAP gates for a given circuit, all possible qubit orderings can be exhaustively searched. However, the efficiency of this approach is limited by the large search space. On the other hand, for two qubits positioned at the locations $i_1$ and $i_2$ ($i_2\geq i_1$), only those qubits that are placed between them need to be considered (i.e., $(i_2 - i_1)!$ permutations instead of all $n!$ permutations) \cite{HirataQIC11}. To further improve runtime, \citeN{HirataQIC11} considered only $(i_2 - i_1)$ permutations for each gate and analyzed only $w$ consecutive gates instead of considering all possible gates as performed by an exhaustive method. Applying the techniques of \cite{HirataQIC11} with $w=10$ on the 8-qubit AQFT$_5$ circuit\footnote{The Quantum Fourier Transform plays a key role in many quantum algorithms. As the number of input qubits grows, QFT needs exponentially smaller phase shifts, which complicates its physical implementation. Therefore, the Approximate Quantum Fourier Transform (AQFT$_m$) was defined by circuits created from QFT
except that all phase shift gates $R_p$ with phase $2\pi/2^p$ are ignored for $m > p$ ($m$ is the approximation parameter). While a QFT of size $n$ requires O($n^2$) gates to implement, \citeN{Cheung:2004} showed that AQFT$_m$, $m=\log_2 n$, with O($n \log_2 n$) gates achieves almost the same accuracy level.} improved hand-optimized results of \cite{Takahashi:2007}. The cost of the AQFT circuit was further optimized by templates introduced in \cite{SaeediQIP11}. Minimizing AQFT circuits is an open challenge.

Key synthesis and optimization algorithms are compared in Table \ref{table:summary}.

\begin{table}[tb]
\tbl{Synthesis and optimization algorithms for reversible circuits.\label{table:summary}}{
\scriptsize
\begin{tabular}{|l|l|l|l|l|}
  \hline
    \sc{Synthesis method }   & \sc{Features} & \sc{Limitations} & \sc{Library} & \sc{Metric} \\
  \hline
  \hline
  \cite{ShendeTCAD03} & Heuristic synthesis  & Circuit dependency & NCT   & QC    \\
                      & Fast  & & & \\
                      & No garbage & & & \\
  \hline
  \cite{PrasadJETC06} & Optimization & Library dependency   & NCT & QC\\
                      & Fast  &  Local optimum only & & \\
                      &  No garbage  & Circuit dependency & & \\
                      & & Windowing strategy& & \\
  \hline
  \cite{GuptaTCAD06} & Heuristic synthesis  & Limited scalability & NCT & QC\\
                      & No garbage &  &  & \\
  \hline
  \cite{HungTCAD06} & Optimal synthesis &   Limited scalability  & NCV & QC\\
  \hline
  \cite{MaslovTODAES07} & Heuristic synthesis & Large runtime & NCT & QC\\
                        & Optimization & Function dependency & & \\
                        &  &Local optimum only &  & \\
                       & & Windowing strategy& & \\
  \hline
  \cite{MaslovTCAD08} & Optimization & Windowing strategy & NCT & Depth\\
                      & Fast  &  Local optimum only  & & \\
  \hline
  \cite{DonaldJETC08} & Heuristic synthesis  & Limited scalability & NCTSFP & QC\\
                      & No garbage &  & & \\
  \hline
  \cite{GrosseTCAD09} & Optimal synthesis & Limited scalability  & NCT & QC\\
  \hline
  \cite{WilleDAC09} & Heuristic synthesis & Numerous ancillae & NCT & QC \\
                      & Fast, scalable & & & \\
                      & Compact circuits & & & \\
  \hline
  \cite{WilleDAC10} & Optimization & Local optimum only & NCT & Ancilla  \\
                    & Fast & Circuit dependency  & & \\
                    & & Windowing strategy& & \\
  \hline
  \cite{ArabzadehASPDAC10} & Optimization & Local optimum only & NCT & QC \\
                      & Fast & Circuit dependency & &\\
                      & & Windowing strategy& & \\
  \hline
  \cite{SaeediJETC10} & Heuristic synthesis & Function dependency & NCT & QC \\
                      & Fast  & & & \\
                      & No garbage &  & & \\

  \hline
  \cite{SaeediQIP11} & Optimization & Local optimum only & Any & Locality \\
                      & Fast  & Circuit dependency & & \\
  \hline
  \cite{HirataQIC11} & Optimization & Local optimum only  & Any & Locality \\
                      & & Circuit dependency & & \\

  \hline
  \cite{MaslovTCAD11} & Optimization & Local optimum only & Any & QC\\
                      & Fast & Circuit dependency &  & \\
  \hline
\end{tabular}}
\end{table}
\section{Benchmarks and Software Tools} \label {sec:bnch}
To analyze the effectiveness of reversible-logic synthesis algorithms, a variety of benchmark functions are available. \citeN{MaslovSite} developed and has been maintaining the \emph{Reversible Logic Synthesis Benchmarks Page} which offers the widely-used benchmark functions for reversible logic and their best-known circuits (as communicated to the maintainer). \texttt{RevLib} introduced in \cite{WilleMVL08} is not limited to best-known circuits and, in addition to some results from \cite{MaslovSite}, includes a variety of sub-optimal circuits. The open-source toolkit \texttt{RevKit} \cite{SFWD:10} includes several utilities and implements algorithms for reversible circuit synthesis. In \cite{WOD:2010}, a programming language was proposed to specify a reversible transformation from which a compiler can generate reversible circuits. A circuit browser, \texttt{RCViewer}, was developed by Scott and Maslov in 2003, and later described in \cite{MaslovSite}. An improved version, \texttt{RCViewer+}, was reported in \cite{RCviewerPlus}. RevKit and RCViewer+ support a number of features --- circuit visualization and cost analysis, equivalence checking, and circuit diagram plotting using \LaTeX\ q-circuit format (http://www.cquic.org/Qcircuit/).

The most common reversible benchmark families are as follows.
\begin{itemize}
    \item [$\bullet$] \textbf{Reversible functions with known optimal circuits} include all 3-input \cite{ShendeTCAD03} and 4-input reversible functions \cite{GolubitskyDAC10}.
        \begin{itemize}
        \vspace{0.1cm}
            \item [$\circ$] \emph{4-bit functions with maximal gate count}: This set introduced by Golubitsky and Maslov \cite{MaslovSite} contains all 4-bit functions whose optimal implementations use 15 gates (largest number possible).
        \vspace{0.1cm}
            \item [$\circ$] \emph{Gray code transforms}: The $N$-bit transform \texttt{GraycodeN} converts binary-coded integers to Gray-coded integers. As this function is GF(2)-linear, an optimal circuit requires only CNOT gates: CNOT($b$,$a$) CNOT($c$,$b$) .... CNOT($z$,$y$) for qubits $a$, $b$, ..., $z$. Several heuristics \cite{GuptaTCAD06,MaslovTODAES07,SaeediJETC10} produce optimal circuits for this family of functions.
        \end{itemize}
    \vspace{0.4cm}
    \item [$\bullet$] \textbf{Arithmetic functions} have applications in quantum algorithms \cite{Childs10}. In conventional circuit design, 32-bit and 64-bit arithmetic circuits are of significant interest because they are used in word-level CPUs. Sophisticated optimizations have been developed for such special cases \cite{Dimitrov:2011}. However, no such \emph{standard sizes} have been established for reversible circuits and applications in quantum computing suggest that such standardization is highly unlikely. Therefore, the design of arithmetic circuits focuses on scalable benchmarks and synthesis algorithms rather than a handful of super-optimized circuits. Another distinction from conventional logic circuits is that (as of 2011) we are unable to motivate studies in reversible implementations of floating-point arithmetic.

        \begin{itemize}
        \vspace{0.1cm}
            \item [$\circ$] \emph{Adders}: The function \texttt{nbitadder} introduced in \cite{Feynman86} has two $n$-bit inputs $A$ and $B$ and one ($n+1$)-bit output $A+B$. Quantum circuits for elementary arithmetic operations are important for the implementation of Shor's factorization algorithm. With one ancilla, a quantum circuit with depth $2n + 4$ and size $9n - 8$ was proposed in \cite{Cuccaro:quant-ph0410184}. A variant method yields a circuit with size $6n + 1$ and depth $6n + 1$. \citeN{Takahashi:2010} proposed a quantum circuit with depth $5n - 3$ and size $7n - 6$ for \texttt{nbitadder} with no ancilla and an O(d($n$))-depth O($n$)-size quantum circuit with O($\frac{n}{d(n)}$) ancillae where $d(n) = \Omega(\log n)$.

        \vspace{0.1cm}
            \item [$\circ$] \emph{Modulo adders}: The function \texttt{modNadder} has $2 \lceil \log_2 N\rceil$ inputs/outputs where for each codeword, the input is a pair of modulo-$N$ numbers $A$ and $B$, while the output is the pair of modulo-$N$ numbers ($A$, $A+B \mod N$). As of 2011, the best results for \texttt{modNadder} functions were obtained by \cite{MaslovTODAES07}

        \vspace{0.1cm}
            \item [$\circ$] \emph{Galois field multipliers}: The Galois field multiplication function \texttt{gfp$\wedge$mmult} \cite{CheungQIC09} has $2m \lceil \log_2 p \rceil$ inputs and $m \lceil \log_2 p \rceil$ outputs. It computes the field product of two GF($p^m$) elements, $a$ and $b$. GF($p^m$) is used in a quantum polynomial-time algorithm that computes discrete logarithm over an elliptic curve group, and it has applications in quantum cryptography. An O($m$)-depth multiplication circuit for GF($2^m$) targeted for an LNN architecture was proposed in \cite{CheungQIC09}.

          \vspace{0.1cm}
          \item [$\circ$] \emph{Divisibility checkers}: The function \texttt{NmodK} has $N$ inputs and a single output. Its output is 1 for those codewords that are divisible by the integer $K$. As of 2011, the best results for \texttt{NmodK} functions were obtained by \cite{MaslovTCAD05}.
        \end{itemize}

        \vspace{0.4cm}
    \item [$\bullet$] \textbf{Hard benchmarks} are mainly proposed to stress-test the existing synthesis algorithms. These functions may be produced from hard benchmarks developed for conventional logic synthesis.
            \begin{itemize}

         \vspace{0.1cm}
               \item [$\circ$] \emph{Hidden weighted-bit function}: The function \texttt{hwbN} has $N$ inputs/outputs where the input codeword is cyclically shifted by the number of ones it has. The conventional HWB function returns the value of the input bit, indexed by the number of ones (mod $n$), and all of its ROBDDs have exponential size \cite{BLSW:1999}. Markov and Maslov showed that \texttt{hwbN} functions can be implemented with a polynomial cost O$(n \log^2 n)$ if a logarithmic number of garbage bits $[\log n]+1$ is available \cite{MaslovSite}. Efficient synthesis with no garbage bits remains open. Known circuits for \texttt{hwb} functions with no ancilla exhibit exponential number of gates. As of 2011, the best results for medium-size \texttt{hwbN} functions with no ancilla were obtained by applying the method of \cite{SaeediJETC10}.

          \vspace{0.1cm}
              \item [$\circ$] \emph{Reversible variants of high-complexity functions}.
                    \begin{itemize}

         \vspace{0.1cm}
                       \item [$\diamond$] \emph{Computation of $N$-th prime}: The function \texttt{nth$\_$primeK$\_$inc} introduced as a reversible benchmark in \cite{MaslovSite} has $K$ inputs/outputs. For an input value $n$, this function returns the $n$-th prime, as long as this prime may be written using at most $K$ bits. The algorithm in \cite{Lagarias1987173} runs in exponential time, and no poly-time circuits or algorithms are known as of 2011. The smallest circuits to date are shown in \cite{SaeediJETC10}. For a simpler problem --- primality testing --- polynomial circuits are proven to exist, but no practical constructions are known as of 2011
                            \cite{Roy2003}.

           \vspace{0.1cm}
                     \item [$\diamond$] \emph{Computation of the matrix permanent}: The function \emph{permanent $N$x$N$} introduced by Maslov as a reversible benchmark has $N^2$ inputs and $\lceil \log(N!) \rceil$ outputs. It computes the permanent of a 0-1 matrix.\footnote{The permanent of an $n \times n$ matrix $A = (a_{i,j})$ is defined as $\operatorname{perm}(A)=\sum_{\sigma\in S_n}\prod_{i=1}^n a_{i,\sigma(i)}$ where $S_n$ is the symmetric group and $\sigma$ is a permutation in $S_n$. For example, for $N=2$ the permanent is $a_{1,1}a_{2,2}+a_{1,2}a_{2,1}$.} There is strong evidence that no polynomial-time non-quantum algorithm exists for this computation \cite{Jerrum:2004}.
                    \end{itemize}
        \end{itemize}
\end{itemize}

Other reversible functions considered as benchmarks \cite{MaslovSite,WilleMVL08,GuptaTCAD06,QECC} include Hamming coding functions and quantum error-correcting codes.

\section{Conclusion and Future Directions} \label {sec:open}
Reversible logic circuits have been studied for at least 30 years \cite{Toffoli80}, with several different motivations in mind --- from low-power computing and bit-twiddles in computer graphics algorithms, to photonic circuits and quantum information processing. Synthesis of reversible logic circuits is typically partitioned into $(i)$ a pre-synthesis optimization that revises the specification, $(ii)$ synthesis per se, $(iii)$ post-synthesis local optimization, and $(iv)$ technology mapping that reflects specific limitations of a given implementation technology.

Despite significant progress in reversible logic synthesis, a number of open challenges remain --- some are in the domain of reversible circuits and others in the broader domain of quantum information processing. In particular, existing reversible synthesis techniques do not perform well on important benchmarks such as arithmetic functions --- they produce circuits that are much larger compared to known solutions.

\begin{itemize}
    \item [$\bullet$] \textbf{Traditional reversible logic synthesis.}
        \begin{itemize}
            \item [$\circ$] Scalable synthesis of general reversible functions targeting different cost models and gate libraries without significant overhead.
            \item [$\circ$] Technology mapping (see Section \ref{sec:picture}) for specific applications, gate libraries, and cost models.
            \item [$\circ$] Optimal (or efficient) synthesis of reversible functions useful in specific applications, e.g., quantum algorithms such as Shor's number-factoring.
            \item [$\circ$] Efficient synthesis algorithms for T-constructible permutations.
        \end{itemize}

    \item [$\bullet$] \textbf{Lower and upper bounds, and worst-case reversible functions.}
        \begin{itemize}
            \item [$\circ$] Sharper lower and upper bounds on the number of elementary gates for reversible functions. Current lower bound is $n2^n / \log n$ \cite{ShendeTCAD03} and upper bound is $8.5n2^n + o(2^n)$ \cite{SaeediJETC10}.
            \item [$\circ$] Lower and upper bounds on the number of elementary gates for T-constructible reversible functions.
            \item [$\circ$] Super-linear lower bounds on the size and depth of NC-circuits (and stabilizer circuits) for specific functions \cite{Aaronson04}.
            \item [$\circ$] The minimal number of CNOT gates required for the implementation of an $n$-qubit Toffoli (or other useful gates such as Fredkin) gate with and without ancillae. Without ancilla, the number of CNOT gates is $\Theta(n^2)$, and $\Theta(n)$ gates are sufficient when at least one ancilla is available \cite{Barenco95}.
        \end{itemize}

    \item [$\bullet$] \textbf{Optimization of quantum circuits.}
        \begin{itemize}
            \item [$\circ$] Synthesis of circuits with provably minimum size (and depth) for stabilizer (or GF(2)-linear) operators \cite{Aaronson04}.
            \item [$\circ$] Small quantum circuits for permutation functions \cite{MaslovTCAD11}.
        \end{itemize}
\end{itemize}

In addition to these rather specific challenges, entirely new concepts and techniques may be discovered for representing reversible functions and synthesizing reversible circuits. Further considerations for future research are summarized below.\\

\textbf{Keeping applications in mind.} Given that reversible logic circuits today are largely motivated by quantum, nano and photonic computing, we note that these novel computing paradigms promise improvements in rather narrow circumstances, while suffering from serious general drawbacks. For example, quantum algorithms are likely to be handicapped by size limitations, as well as quantum noise and decoherence, but offer polynomial-time algorithms for certain problems where conventional computers currently spend exponential time \cite{Nielsen00}. Therefore, it does not necessarily make sense to study reversible versions of every conventional circuit. Aside from trying to stress-test logic synthesis tools, specific reversible circuits must be motivated by applications. For example, reversible adders and modular multipliers have been motivated by Shor's quantum number-factoring algorithm \cite{Beckman96,Meter05,MarkovQIC2012,Markov2013} which leverages unique properties of quantum circuits. 

        \vspace{0.1cm}
\textbf{Sequential reversible computation} has been studied as early as in the 1980s in \cite{Toffoli80}, but research on this topic is still lacking sufficient motivation. In conventional circuits, sequential elements are clocked, but reversible clocking has not been considered (and may not make sense, since this is not a logic operation). This undermines considerations of low power for sequential reversible computation, as clocking and clocked elements consume a large fraction of energy used by CMOS circuits. Most uses of reversible transformations in cryptography, DSP and computer graphics are combinational in nature. Most quantum computers use stationary qubits and apply gates to these qubits, unlike CMOS circuits where gates are stationary and signals traverse the circuits. In the context of stationary qubits, quantum (and reversible) circuits already have significant sequential semantics \cite{Morita08}, and there is no need for dedicated sequential elements. Algorithms for the design, analysis and verification of Quantum Finite Automata \cite{Moore:2000} may be of some interest, if sufficiently motivated by applications.

        \vspace{0.1cm}
\textbf{Verification of reversible circuits} is important because circuit optimization algorithms and software tools have become so complex that subtle bugs are very likely. Fortunately, equivalence-checking techniques for conventional combinational circuits can be applied by converting CNOT gates to XORs, Toffoli gates into ANDs and XORs, and so on (Fig. \ref{fig:Fig1}a). Additional efficiency improvements can be obtained by exploiting reversibility as illustrated in \cite{Wille:2009,Yamashita10}. Verification of non-Boolean quantum circuits is more challenging and, in general, appears as hard as quantum simulation \cite{Viamontes:2007,Yamashita10}.

        \vspace{0.1cm}
\textbf{Circuit test} is vital to check if a circuit works as expected. Efficient test is critical for mass-production facilities, whereas laboratory experiments emphasize precision. Test techniques are sensitive to dominant fault types, but fortunately CMOS circuits can be tested reasonably well assuming only stuck-at fault models \cite{Bushnell:2000}. Given that CMOS is not the dominant technology for reversible circuits, the use of stuck-at fault models in this context may be unjustified, or at least requires explicit justification. An example reversible fault model is given in \cite{Polian:2005}, where circuit test is performed for missing gates.\footnote{In ion-trap and NMR quantum computers, gates are effected by RF pulses. If the wavelength drifts too far from the desired value, the gate will not be applied. This situation illustrates \emph{stuck-at-0 faults} on controls of CNOT and Toffoli gates, but stuck-at faults on bit lines would imply the loss of reversibility.} In the case of quantum circuits, test is particularly complicated because measurements produce nondeterministic results. Therefore, quantum-computing experiments are typically verified using tomography \cite{Altepeter2005105}, i.e., plotting an entire distribution of possible outcomes.

        \vspace{0.1cm}
\textbf{Error-detection and fault-tolerance techniques} are motivated for circuit technologies that are likely to experience transient faults, which is the case with quantum circuits. Like circuit test, these techniques are heavily dependent on fault models and measurement, and naive attempts to model quantum faults and error-detection by Boolean techniques lead to nonsensical results. Fault-tolerant quantum computing is an extensively developed branch of quantum information processing, and its basics are introduced in standard textbooks \cite{Nielsen00}.

        \vspace{0.1cm}
\textbf{Quantum-logic synthesis} deals with general unitary matrices and is more challenging than reversible-logic synthesis. As of 2011, the most compact circuit constructions use $\frac{23}{48} 4^n  - \frac {3}{2} 2^n  + \frac{4}{3}$ CNOTs \cite{Shende06,Mottonen06} and $\frac{1}{2} 4^n  + \frac{1}{2} 2^n  - n - 1$ one-qubit gates \cite{Bergholm:2004}. The sharpest lower bound on the number of CNOT gates is $\left\lceil \frac{1}{4}(4^n  - 3n - 1) \right\rceil$ \cite{shende-2004}. Different trade-offs between the number of one-qubit gates and CNOTs are explored in \cite{SaeediQIC11}. Future research directions include simultaneous reduction of CNOT and one-qubit gates \cite{Shende09}, sharper lower bounds on the number of one-qubit and CNOT gates, and consideration of circuit depth, perhaps with ancillae.

        \vspace{0.1cm}
\textbf{Physical layout and optimization of quantum circuits} are crucial to map circuit qubits into physical qubits. Currently, the layout of quantum circuits is hand-optimized when preparing laboratory experiments, but automated techniques are required to systematically accomplish this task as the capacity of quantum computers increases. In \cite{MaslovTCAD08placement}, the authors proposed a heuristic for the placement problem by optimizing qubit-to-qubit interaction and showed that the problem of mapping circuit qubits to physical qubits is NP-complete.

        \vspace{0.1cm}
\textbf{Physical implementation of reversible circuits} using switching devices with limited or no gain may generate new applications. Aside from quantum circuits, interesting examples include implementations in CMOS powered by circuit inputs \cite{Desoete02,DeVos:2010,Skoneczny} as well
as photonic circuits \cite{Politi09,Gao2010}.

        \vspace{0.1cm}
\textbf{Design and verification tools} for reversible and quantum circuits have been developed and reported by a number of groups, but in most cases they are point tools built to demonstrate specific algorithms. In contrast, conventional circuit-design environments employ long chains of inter-operating software tools. Such powerful software may be necessary to scale reversible and quantum circuit design beyond its current limitations \cite{Svore:2006,WilleBook}. On the other hand, there is danger of developing CAD tools that are not fully motivated by applications.

\begin{acks}
We thank Dr. Dmitri Maslov, Dr. Vivek Shende, H\'{e}ctor J. Garc\'{i}a, Prof. John P. Hayes, and Dr. Smita Krishnaswamy for proofreading early versions of the manuscript and for helpful discussions.
\end{acks}


\begin{thebibliography}{}

\bibitem[\protect\citeauthoryear{Aaronson and Gottesman}{Aaronson and
  Gottesman}{2004}]{Aaronson04}
{\sc Aaronson, S.} {\sc and} {\sc Gottesman, D.} 2004.
\newblock Improved simulation of stabilizer circuits.
\newblock {\em Phys. Rev. A\/}~{\em 70}, 052328, arXiv:quant--ph/0406196v5.

\bibitem[\protect\citeauthoryear{Altepeter, Jeffrey, and Kwiat}{Altepeter
  et~al\mbox{.}}{2005}]{Altepeter2005105}
{\sc Altepeter, J.}, {\sc Jeffrey, E.}, {\sc and} {\sc Kwiat, P.} 2005.
\newblock Photonic state tomography.
\newblock Advances in Atomic, Molecular, and Optical Physics Series, vol.~52.
  Academic Press, 105--159.

\bibitem[\protect\citeauthoryear{Arabzadeh and Saeedi}{Arabzadeh and
  Saeedi}{2011}]{RCviewerPlus}
{\sc Arabzadeh, M.} {\sc and} {\sc Saeedi, M.} 2011.
\newblock {RCviewer+, A} viewer/analyzer for reversible and quantum circuits,
  version 1.88.
\newblock {\em http://ceit.aut.ac.ir/QDA/RCV.htm\/}.

\bibitem[\protect\citeauthoryear{Arabzadeh, Saeedi, and {Saheb
  Zamani}}{Arabzadeh et~al\mbox{.}}{2010}]{ArabzadehASPDAC10}
{\sc Arabzadeh, M.}, {\sc Saeedi, M.}, {\sc and} {\sc {Saheb Zamani}, M.} 2010.
\newblock Rule-based optimization of reversible circuits.
\newblock {\em Asia and South Pacific Design Autom. Conf.\/}, 849--854.

\bibitem[\protect\citeauthoryear{{Asano} and {Ishii}}{{Asano} and
  {Ishii}}{2005}]{Asano05}
{\sc {Asano}, M.} {\sc and} {\sc {Ishii}, C.} 2005.
\newblock New structural quantum circuit simulating a {Toffoli} gate.
\newblock {\em arXiv:quant-ph/0512016\/}.

\bibitem[\protect\citeauthoryear{Bacon and van Dam}{Bacon and van
  Dam}{2010}]{Bacon:2010}
{\sc Bacon, D.} {\sc and} {\sc van Dam, W.} 2010.
\newblock Recent progress in quantum algorithms.
\newblock {\em Commun. ACM\/}~{\em 53}, 84--93.

\bibitem[\protect\citeauthoryear{Barenco, Bennett, Cleve, DiVincenzo, Margolus,
  Shor, Sleator, Smolin, and Weinfurter}{Barenco
  et~al\mbox{.}}{1995}]{Barenco95}
{\sc Barenco, A.}, {\sc Bennett, C.}, {\sc Cleve, R.}, {\sc DiVincenzo, D.},
  {\sc Margolus, N.}, {\sc Shor, P.}, {\sc Sleator, T.}, {\sc Smolin, J.}, {\sc
  and} {\sc Weinfurter, H.} 1995.
\newblock Elementary gates for quantum computation.
\newblock {\em Phys. Rev. A\/}~{\em 52}, 3457--3467.

\bibitem[\protect\citeauthoryear{Beckman, Chari, Devabhaktuni, and
  Preskill}{Beckman et~al\mbox{.}}{1996}]{Beckman96}
{\sc Beckman, D.}, {\sc Chari, A.~N.}, {\sc Devabhaktuni, S.}, {\sc and} {\sc
  Preskill, J.} 1996.
\newblock Efficient networks for quantum factoring.
\newblock {\em Phys. Rev. A\/}~{\em 54,\/}~2, 1034--1063.

\bibitem[\protect\citeauthoryear{Bennett}{Bennett}{1973}]{Bennett73}
{\sc Bennett, C.~H.} 1973.
\newblock Logical reversibility of computation.
\newblock {\em IBM J. Resear. Deve.\/}~{\em 17,\/}~6, 525--532.

\bibitem[\protect\citeauthoryear{Bergholm, Vartiainen, M\"{o}tt\"{o}nen, and
  Salomaa}{Bergholm et~al\mbox{.}}{2005}]{Bergholm:2004}
{\sc Bergholm, V.}, {\sc Vartiainen, J.~J.}, {\sc M\"{o}tt\"{o}nen, M.}, {\sc
  and} {\sc Salomaa, M.~M.} 2005.
\newblock Quantum circuits with uniformly controlled one-qubit gates.
\newblock {\em Phys. Rev. A\/}~{\em 71}, 052330.

\bibitem[\protect\citeauthoryear{Bollig, L\"obbing, Sauerhoff, and
  Wegener}{Bollig et~al\mbox{.}}{1999}]{BLSW:1999}
{\sc Bollig, B.}, {\sc L\"obbing, M.}, {\sc Sauerhoff, M.}, {\sc and} {\sc
  Wegener, I.} 1999.
\newblock {On The Complexity of the Hidden Weighted Bit Function for Various
  {BDD} Models}.
\newblock {\em Informatique Theorique et Applications\/}~{\em 33,\/}~2,
  103--116.

\bibitem[\protect\citeauthoryear{Bryant}{Bryant}{1986}]{Bryant86}
{\sc Bryant, R.} 1986.
\newblock Graph-based algorithms for boolean function manipulation.
\newblock {\em IEEE Trans. Comput.\/}~{\em 35,\/}~8, 677--691.

\bibitem[\protect\citeauthoryear{Bushnell and Agrawal}{Bushnell and
  Agrawal}{2000}]{Bushnell:2000}
{\sc Bushnell, M.} {\sc and} {\sc Agrawal, V.} 2000.
\newblock {\em Essentials of Electronic Testing for Digital, Memory and
  Mixed-Signal VLSI Circuits}.
\newblock Kluwer, USA.

\bibitem[\protect\citeauthoryear{Cheung}{Cheung}{2004}]{Cheung:2004}
{\sc Cheung, D.} 2004.
\newblock Improved bounds for the approximate {QFT}.
\newblock {\em Int'l Symp. on Inf. and Communication Technologies\/}, 1--6.

\bibitem[\protect\citeauthoryear{Cheung, Maslov, Mathew, and Pradhan}{Cheung
  et~al\mbox{.}}{2009}]{CheungQIC09}
{\sc Cheung, D.}, {\sc Maslov, D.}, {\sc Mathew, J.}, {\sc and} {\sc Pradhan,
  D.~K.} 2009.
\newblock On the design and optimization of a quantum polynomial-time attack on
  elliptic curve cryptography.
\newblock {\em Quant. Inf. Comput.\/}~{\em 9,\/}~7\&8, 610--621.

\bibitem[\protect\citeauthoryear{Cheung, Maslov, and Severini.}{Cheung
  et~al\mbox{.}}{2007}]{Cheung07}
{\sc Cheung, D.}, {\sc Maslov, D.}, {\sc and} {\sc Severini., S.} 2007.
\newblock Translation techniques between quantum circuit architectures.
\newblock {\em Workshop on Quant. Inf. Proc.\/}.

\bibitem[\protect\citeauthoryear{Childs and van Dam}{Childs and van
  Dam}{2010}]{Childs10}
{\sc Childs, A.~M.} {\sc and} {\sc van Dam, W.} 2010.
\newblock Quantum algorithms for algebraic problems.
\newblock {\em Rev. Mod. Phys.\/}~{\em 82,\/}~1, 1--52.

\bibitem[\protect\citeauthoryear{{Choi} and {Van Meter}}{{Choi} and {Van
  Meter}}{2008}]{Choi08}
{\sc {Choi}, B.} {\sc and} {\sc {Van Meter}, R.} 2008.
\newblock Effects of interaction distance on quantum addition circuits.
\newblock {\em quant-ph/0809.4317\/}.

\bibitem[\protect\citeauthoryear{Cuccaro, Draper, Kutin, and Moulton}{Cuccaro
  et~al\mbox{.}}{2005}]{Cuccaro:quant-ph0410184}
{\sc Cuccaro, S.~A.}, {\sc Draper, T.~G.}, {\sc Kutin, S.~A.}, {\sc and} {\sc
  Moulton, D.~P.} 2005.
\newblock A new quantum ripple-carry addition circuit.
\newblock {\em Workshop on Quant. Inf. Proc.\/}.

\bibitem[\protect\citeauthoryear{Dally and Towles}{Dally and
  Towles}{2003}]{Dally:2003}
{\sc Dally, W.} {\sc and} {\sc Towles, B.} 2003.
\newblock {\em Principles and Practices of Interconnection Networks}.
\newblock Morgan Kaufmann Publishers Inc., USA.

\bibitem[\protect\citeauthoryear{De~Vos}{De~Vos}{2010a}]{DeVos:2010}
{\sc De~Vos, A.} 2010a.
\newblock Reversible computer hardware.
\newblock {\em Electron. Notes Theor. Comput. Sci.\/}~{\em 253,\/}~6, 17--22.

\bibitem[\protect\citeauthoryear{De~Vos}{De~Vos}{2010b}]{DeVosBook}
{\sc De~Vos, A.} 2010b.
\newblock {\em Reversible Computing}.
\newblock Wiley–VCH.

\bibitem[\protect\citeauthoryear{{De Vos}, Raa, and Storme}{{De Vos}
  et~al\mbox{.}}{2002}]{DeVos2002}
{\sc {De Vos}, A.}, {\sc Raa, B.}, {\sc and} {\sc Storme, L.} 2002.
\newblock Generating the group of reversible logic gates.
\newblock {\em J. of Phys. A\/}~{\em 35,\/}~33, 7063.

\bibitem[\protect\citeauthoryear{Desoete and De~Vos}{Desoete and
  De~Vos}{2002}]{Desoete02}
{\sc Desoete, B.} {\sc and} {\sc De~Vos, A.} 2002.
\newblock A reversible carry-look-ahead adder using control gates.
\newblock {\em Integr. VLSI J.\/}~{\em 33,\/}~1, 89--104.

\bibitem[\protect\citeauthoryear{Dimitrov, Jarvinen, and Adikari}{Dimitrov
  et~al\mbox{.}}{2011}]{Dimitrov:2011}
{\sc Dimitrov, V.~S.}, {\sc Jarvinen, K.~U.}, {\sc and} {\sc Adikari, J.} 2011.
\newblock Area-efficient multipliers based on multiple-radix representations.
\newblock {\em IEEE Trans. Comput.\/}~{\em 60,\/}~2, 189--201.

\bibitem[\protect\citeauthoryear{Donald and Jha}{Donald and
  Jha}{2008}]{DonaldJETC08}
{\sc Donald, J.} {\sc and} {\sc Jha, N.~K.} 2008.
\newblock Reversible logic synthesis with {Fredkin} and {Peres} gates.
\newblock {\em J. Emerg. Technol. Comput. Sys.\/}~{\em 4,\/}~1, 2:1--2:19.

\bibitem[\protect\citeauthoryear{Dou\ifmmode~\mbox{\c{c}}\else \c{c}\fi{}ot,
  Ioffe, and Vidal}{Dou\ifmmode~\mbox{\c{c}}\else \c{c}\fi{}ot
  et~al\mbox{.}}{2004}]{PhysRevB.69.214501}
{\sc Dou\ifmmode~\mbox{\c{c}}\else \c{c}\fi{}ot, B.}, {\sc Ioffe, L.~B.}, {\sc
  and} {\sc Vidal, J.} 2004.
\newblock Discrete non-{Abelian} gauge theories in {Josephson-junction} arrays
  and quantum computation.
\newblock {\em Phys. Rev. B\/}~{\em 69,\/}~21, 214501.

\bibitem[\protect\citeauthoryear{Egner, P\"{u}schel, and Beth}{Egner
  et~al\mbox{.}}{1997}]{Egner:1997}
{\sc Egner, S.}, {\sc P\"{u}schel, M.}, {\sc and} {\sc Beth, T.} 1997.
\newblock Decomposing a permutation into a conjugated tensor product.
\newblock {\em Int'l Symp. on Symbolic and Algebraic Computation\/}, 101--108.

\bibitem[\protect\citeauthoryear{Fazel, Thornton, and Rice}{Fazel
  et~al\mbox{.}}{2007}]{FTR:2007}
{\sc Fazel, K.}, {\sc Thornton, M.}, {\sc and} {\sc Rice, J.} 2007.
\newblock {ESOP-based Toffoli Gate Cascade Generation}.
\newblock {\em Proceedings of the IEEE Pacific Rim Conference on
  Communications\/}, 206--209.

\bibitem[\protect\citeauthoryear{Fei, Jiang-Feng, Ming-Jun, Xian-Yi, Rong-Dian,
  and Ji-Hui}{Fei et~al\mbox{.}}{2002}]{Fei02}
{\sc Fei, X.}, {\sc Jiang-Feng, D.}, {\sc Ming-Jun, S.}, {\sc Xian-Yi, Z.},
  {\sc Rong-Dian, H.}, {\sc and} {\sc Ji-Hui, W.} 2002.
\newblock Realization of the {Fredkin} gate by three transition pulses in a
  nuclear magnetic resonance quantum information processor.
\newblock {\em Chinese Phys. Lett.\/}~{\em 19,\/}~8, 1048.

\bibitem[\protect\citeauthoryear{Feynman}{Feynman}{1986}]{Feynman86}
{\sc Feynman, R.} 1986.
\newblock Quantum mechanical computers.
\newblock {\em Found. Phys.\/}~{\em 16,\/}~6, 507--531.

\bibitem[\protect\citeauthoryear{Fowler, Devitt, and Hollenberg}{Fowler
  et~al\mbox{.}}{2004a}]{FDH:2004}
{\sc Fowler, A.~G.}, {\sc Devitt, S.~J.}, {\sc and} {\sc Hollenberg, L. C.~L.}
  2004a.
\newblock Implementation of {Shor's} algorithm on a linear nearest neighbour
  qubit array.
\newblock {\em Quant. Inf. Comput.\/}~{\em 4}, 237--245.

\bibitem[\protect\citeauthoryear{Fowler, Hill, and Hollenberg}{Fowler
  et~al\mbox{.}}{2004b}]{Fowler}
{\sc Fowler, A.~G.}, {\sc Hill, C.~D.}, {\sc and} {\sc Hollenberg, L. C.~L.}
  2004b.
\newblock Quantum error correction on linear nearest neighbor qubit arrays.
\newblock {\em Phys. Rev. A\/}~{\em 69,\/}~4, 042314.1--042314.4.

\bibitem[\protect\citeauthoryear{Fredkin and Toffoli}{Fredkin and
  Toffoli}{1982}]{Fredkin82}
{\sc Fredkin, E.~F.} {\sc and} {\sc Toffoli, T.} 1982.
\newblock Conservative logic.
\newblock {\em Int. J. Theor. Phys.\/}~{\em 21,\/}~3/4, 219--253.

\bibitem[\protect\citeauthoryear{Gao, Xu, Yao, G\"uhne, Cabello, Lu, Peng,
  Chen, and Pan}{Gao et~al\mbox{.}}{2010}]{Gao2010}
{\sc Gao, W.-B.}, {\sc Xu, P.}, {\sc Yao, X.-C.}, {\sc G\"uhne, O.}, {\sc
  Cabello, A.}, {\sc Lu, C.-Y.}, {\sc Peng, C.-Z.}, {\sc Chen, Z.-B.}, {\sc
  and} {\sc Pan, J.-W.} 2010.
\newblock Experimental realization of a controlled-{NOT} gate with four-photon
  six-qubit cluster states.
\newblock {\em Phys. Rev. Lett.\/}~{\em 104,\/}~2, 020501.

\bibitem[\protect\citeauthoryear{Gl\"{u}ck and Kawabe}{Gl\"{u}ck and
  Kawabe}{2005}]{Gluck:2005}
{\sc Gl\"{u}ck, R.} {\sc and} {\sc Kawabe, M.} 2005.
\newblock A method for automatic program inversion based on lr(0) parsing.
\newblock {\em Fundam. Inf.\/}~{\em 66,\/}~4, 367--395.

\bibitem[\protect\citeauthoryear{Golubitsky, Falconer, and Maslov}{Golubitsky
  et~al\mbox{.}}{2010}]{GolubitskyDAC10}
{\sc Golubitsky, O.}, {\sc Falconer, S.~M.}, {\sc and} {\sc Maslov, D.} 2010.
\newblock Synthesis of the optimal 4-bit reversible circuits.
\newblock {\em Design Autom. Conf.\/}, 653--656.

\bibitem[\protect\citeauthoryear{Golubitsky and Maslov}{Golubitsky and
  Maslov}{2011}]{Golubitsky11}
{\sc Golubitsky, O.} {\sc and} {\sc Maslov, D.} 2011.
\newblock A study of optimal 4-bit reversible toffoli circuits and their
  synthesis.
\newblock {\em arXiv:1103.2686\/}.

\bibitem[\protect\citeauthoryear{Grassl}{Grassl}{2003}]{QECC}
{\sc Grassl, M.} 2003.
\newblock Circuits for quantum error-correcting codes.
\newblock {\em http://iaks-www.ira.uka.de/home/grassl/QECC/index.html\/}.

\bibitem[\protect\citeauthoryear{Grosse, Wille, Dueck, and Drechsler}{Grosse
  et~al\mbox{.}}{2009}]{GrosseTCAD09}
{\sc Grosse, D.}, {\sc Wille, R.}, {\sc Dueck, G.}, {\sc and} {\sc Drechsler,
  R.} 2009.
\newblock Exact multiple-control {Toffoli} network synthesis with {SAT}
  techniques.
\newblock {\em IEEE Trans. CAD\/}~{\em 28,\/}~5, 703--715.

\bibitem[\protect\citeauthoryear{Gupta, Agrawal, and Jha}{Gupta
  et~al\mbox{.}}{2006}]{GuptaTCAD06}
{\sc Gupta, P.}, {\sc Agrawal, A.}, {\sc and} {\sc Jha, N.} 2006.
\newblock An algorithm for synthesis of reversible logic circuits.
\newblock {\em IEEE Trans. CAD\/}~{\em 25,\/}~11, 2317--2330.

\bibitem[\protect\citeauthoryear{Hachtel and Somenzi}{Hachtel and
  Somenzi}{2000}]{Hachtel:2000}
{\sc Hachtel, G.~D.} {\sc and} {\sc Somenzi, F.} 2000.
\newblock {\em Logic Synthesis and Verification Algorithms}.
\newblock Kluwer.

\bibitem[\protect\citeauthoryear{H\"{a}ffner, H\"{a}nsel, Roos, Benhelm,
  al~kar, Chwalla, K\"{o}rber, Rapol, Riebe, Schmidt, Becher, G\"{u}hne,
  D\"{u}r, and Blatt}{H\"{a}ffner et~al\mbox{.}}{2005}]{TrappedIon}
{\sc H\"{a}ffner, H.}, {\sc H\"{a}nsel, W.}, {\sc Roos, C.~F.}, {\sc Benhelm,
  J.}, {\sc al~kar, D.~C.}, {\sc Chwalla, M.}, {\sc K\"{o}rber, T.}, {\sc
  Rapol, U.~D.}, {\sc Riebe, M.}, {\sc Schmidt, P.~O.}, {\sc Becher, C.}, {\sc
  G\"{u}hne, O.}, {\sc D\"{u}r, W.}, {\sc and} {\sc Blatt, R.} 2005.
\newblock Scalable multiparticle entanglement of trapped ions.
\newblock {\em Nature\/}~{\em 438}, 643--646.

\bibitem[\protect\citeauthoryear{Hilewitz and Lee}{Hilewitz and
  Lee}{2008}]{Hilewitz08}
{\sc Hilewitz, Y.} {\sc and} {\sc Lee, R.~B.} 2008.
\newblock Fast bit gather, bit scatter and bit permutation instructions for
  commodity microprocessors.
\newblock {\em J. of Signal Proc. Sys.\/}~{\em 53}, 145--169.

\bibitem[\protect\citeauthoryear{Hirata, Nakanishi, Yamashita, and
  Nakashima}{Hirata et~al\mbox{.}}{2011}]{HirataQIC11}
{\sc Hirata, Y.}, {\sc Nakanishi, M.}, {\sc Yamashita, S.}, {\sc and} {\sc
  Nakashima, Y.} 2011.
\newblock An efficient conversion of quantum circuits to a linear nearest
  neighbor architecture.
\newblock {\em Quant. Inf. Comput.\/}~{\em 11,\/}~1--2, 0142--0166.

\bibitem[\protect\citeauthoryear{Hung, Song, Yang, Yang, and Perkowski}{Hung
  et~al\mbox{.}}{2006}]{HungTCAD06}
{\sc Hung, W.}, {\sc Song, X.}, {\sc Yang, G.}, {\sc Yang, J.}, {\sc and} {\sc
  Perkowski, M.} 2006.
\newblock Optimal synthesis of multiple output {Boolean} functions using a set
  of quantum gates by symbolic reachability analysis.
\newblock {\em IEEE Trans. CAD\/}~{\em 25,\/}~9, 1652--1663.

\bibitem[\protect\citeauthoryear{Iwama, Kambayashi, and Yamashita}{Iwama
  et~al\mbox{.}}{2002}]{IwamaDAC02}
{\sc Iwama, K.}, {\sc Kambayashi, Y.}, {\sc and} {\sc Yamashita, S.} 2002.
\newblock Transformation rules for designing {CNOT}-based quantum circuits.
\newblock {\em Design Autom. Conf.\/}, 419--424.

\bibitem[\protect\citeauthoryear{Jerrum, Sinclair, and Vigoda}{Jerrum
  et~al\mbox{.}}{2004}]{Jerrum:2004}
{\sc Jerrum, M.}, {\sc Sinclair, A.}, {\sc and} {\sc Vigoda, E.} 2004.
\newblock A polynomial-time approximation algorithm for the permanent of a
  matrix with nonnegative entries.
\newblock {\em J. ACM\/}~{\em 51,\/}~4, 671--697.

\bibitem[\protect\citeauthoryear{Kane}{Kane}{1998}]{Kane}
{\sc Kane, B.} 1998.
\newblock A silicon-based nuclear spin quantum computer.
\newblock {\em Nature\/}~{\em 393}, 133--137.

\bibitem[\protect\citeauthoryear{Kerntopf}{Kerntopf}{2004}]{KerntopfDAC04}
{\sc Kerntopf, P.} 2004.
\newblock A new heuristic algorithm for reversible logic synthesis.
\newblock {\em Design Autom. Conf.\/}, 834--837.

\bibitem[\protect\citeauthoryear{Kim, Ziesler, and Papaefthymiou}{Kim
  et~al\mbox{.}}{2005}]{Kim:2005}
{\sc Kim, S.}, {\sc Ziesler, C.~H.}, {\sc and} {\sc Papaefthymiou, M.~C.} 2005.
\newblock Charge-recovery computing on silicon.
\newblock {\em IEEE Trans. Comput.\/}~{\em 54,\/}~6, 651--659.

\bibitem[\protect\citeauthoryear{Korf}{Korf}{1999}]{Korf99}
{\sc Korf, R.} 1999.
\newblock {\em Artificial intelligence search algorithms}.
\newblock in Algorithms Theory Computation Handbook, CRC Press.

\bibitem[\protect\citeauthoryear{Kutin, Moulton, and Smithline}{Kutin
  et~al\mbox{.}}{2007}]{Kutin07}
{\sc Kutin, S.}, {\sc Moulton, D.}, {\sc and} {\sc Smithline, L.} 2007.
\newblock Computation at a distance.
\newblock {\em Chicago J. of Theor. Comput. Sci.\/}.

\bibitem[\protect\citeauthoryear{Kutin}{Kutin}{2007}]{Kutin:2007}
{\sc Kutin, S.~A.} 2007.
\newblock {Shor's} algorithm on a nearest-neighbor machine.
\newblock {\em Asian Conf. on Quant. Inf. Sci.\/}.

\bibitem[\protect\citeauthoryear{Laforest, Simon, Boileau, Baugh, Ditty, and
  Laflamme}{Laforest et~al\mbox{.}}{2007}]{NMR}
{\sc Laforest, M.}, {\sc Simon, D.}, {\sc Boileau, J.-C.}, {\sc Baugh, J.},
  {\sc Ditty, M.}, {\sc and} {\sc Laflamme, R.} 2007.
\newblock Using error correction to determine the noise model.
\newblock {\em Phys. Rev. A\/}~{\em 75,\/}~1, 133--137.

\bibitem[\protect\citeauthoryear{Lagarias and Odlyzko}{Lagarias and
  Odlyzko}{1987}]{Lagarias1987173}
{\sc Lagarias, J.~C.} {\sc and} {\sc Odlyzko, A.~M.} 1987.
\newblock Computing [pi](x): An analytic method.
\newblock {\em J. of Algorithms\/}~{\em 8,\/}~2, 173--191.

\bibitem[\protect\citeauthoryear{Landauer}{Landauer}{1961}]{Landauer61}
{\sc Landauer, R.} 1961.
\newblock Irreversibility and heat generation in the computing process.
\newblock {\em IBM J. Resear. Deve.\/}~{\em 5}, 183--191.

\bibitem[\protect\citeauthoryear{Lee, Lee, Kim, Lee, Biamonte, and
  Perkowski}{Lee et~al\mbox{.}}{2006}]{Lee06}
{\sc Lee, S.}, {\sc Lee, S.}, {\sc Kim, T.}, {\sc Lee, J.~S.}, {\sc Biamonte,
  J.}, {\sc and} {\sc Perkowski, M.} 2006.
\newblock The cost of quantum gate primitives.
\newblock {\em J. of Multiple-Valued Logic and Soft Comput.\/}~{\em
  12,\/}~5--6.

\bibitem[\protect\citeauthoryear{Markov and Roy}{Markov and
  Roy}{2003}]{Roy2003}
{\sc Markov, I.~L.} {\sc and} {\sc Roy, J.~A.} 2003.
\newblock On sub-optimality and scalability of logic synthesis tools.
\newblock {\em Int'l Workshop on Logic Synth.\/}.

\bibitem[\protect\citeauthoryear{Markov and Saeedi}{Markov and
  Saeedi}{2012}]{MarkovQIC2012}
{\sc Markov, I.~L.} {\sc and} {\sc Saeedi, M.} 2012.
\newblock Constant-optimized quantum circuits for modular multiplication and
  exponentiation.
\newblock {\em Quantum Info. Comput.\/}~{\em 12,\/}~5-6, 361--394.

\bibitem[\protect\citeauthoryear{Markov and Saeedi}{Markov and
  Saeedi}{2013}]{Markov2013}
{\sc Markov, I.~L.} {\sc and} {\sc Saeedi, M.} 2013.
\newblock Faster quantum number factoring via circuit synthesis.
\newblock {\em Phys. Rev. A\/}~{\em 87}, 012310.

\bibitem[\protect\citeauthoryear{Maslov}{Maslov}{2007}]{Maslov07}
{\sc Maslov, D.} 2007.
\newblock Linear depth stabilizer and quantum {Fourier} transformation circuits
  with no auxiliary qubits in finite neighbor quantum architectures.
\newblock {\em Phys. Rev. A\/}~{\em 76}.

\bibitem[\protect\citeauthoryear{Maslov}{Maslov}{2011}]{MaslovSite}
{\sc Maslov, D.} Feb 2011.
\newblock Reversible logic synthesis benchmarks page.
\newblock {\em http://www.cs.uvic.ca/\textasciitilde dmaslov/\/}.

\bibitem[\protect\citeauthoryear{Maslov and Dueck}{Maslov and
  Dueck}{2003}]{MaslovIEE03}
{\sc Maslov, D.} {\sc and} {\sc Dueck, G.} 2003.
\newblock Improved quantum cost for $n$-bit {Toffoli} gates.
\newblock {\em Electronics Letters\/}~{\em 39,\/}~25, 1790--1791.

\bibitem[\protect\citeauthoryear{Maslov and Dueck}{Maslov and
  Dueck}{2004}]{MaslovTCAD04}
{\sc Maslov, D.} {\sc and} {\sc Dueck, G.} 2004.
\newblock Reversible cascades with minimal garbage.
\newblock {\em IEEE Trans. CAD\/}~{\em 23,\/}~11, 1497--1509.

\bibitem[\protect\citeauthoryear{Maslov, Dueck, and Miller}{Maslov
  et~al\mbox{.}}{2005a}]{MaslovTVLSI05}
{\sc Maslov, D.}, {\sc Dueck, G.}, {\sc and} {\sc Miller, D.} 2005a.
\newblock Synthesis of {Fredkin-Toffoli} reversible networks.
\newblock {\em IEEE Trans. VLSI\/}~{\em 13,\/}~6, 765--769.

\bibitem[\protect\citeauthoryear{Maslov, Dueck, and Miller}{Maslov
  et~al\mbox{.}}{2005b}]{MaslovTCAD05}
{\sc Maslov, D.}, {\sc Dueck, G.}, {\sc and} {\sc Miller, D.} 2005b.
\newblock {Toffoli} network synthesis with templates.
\newblock {\em IEEE Trans. CAD\/}~{\em 24,\/}~6, 807--817.

\bibitem[\protect\citeauthoryear{Maslov, Dueck, Miller, and Negrevergne}{Maslov
  et~al\mbox{.}}{2008a}]{MaslovTCAD08}
{\sc Maslov, D.}, {\sc Dueck, G.}, {\sc Miller, D.}, {\sc and} {\sc
  Negrevergne, C.} 2008a.
\newblock Quantum circuit simplification and level compaction.
\newblock {\em IEEE Trans. CAD\/}~{\em 27,\/}~3, 436--444.

\bibitem[\protect\citeauthoryear{Maslov, Dueck, and Miller}{Maslov
  et~al\mbox{.}}{2007}]{MaslovTODAES07}
{\sc Maslov, D.}, {\sc Dueck, G.~W.}, {\sc and} {\sc Miller, D.~M.} 2007.
\newblock Techniques for the synthesis of reversible {Toffoli} networks.
\newblock {\em ACM Trans. Des. Autom. Electron. Sys.\/}~{\em 12,\/}~4,
  42:1--42:28.

\bibitem[\protect\citeauthoryear{Maslov, Falconer, and Mosca}{Maslov
  et~al\mbox{.}}{2008b}]{MaslovTCAD08placement}
{\sc Maslov, D.}, {\sc Falconer, S.~M.}, {\sc and} {\sc Mosca, M.} 2008b.
\newblock Quantum circuit placement.
\newblock {\em IEEE Trans. CAD\/}~{\em 27,\/}~4, 752--763.

\bibitem[\protect\citeauthoryear{Maslov and Saeedi}{Maslov and
  Saeedi}{2011}]{MaslovTCAD11}
{\sc Maslov, D.} {\sc and} {\sc Saeedi, M.} 2011.
\newblock Reversible circuit optimization via leaving the {Boolean} domain.
\newblock {\em IEEE Trans. CAD\/}~{\em 30,\/}~6, 806 -- 816.

\bibitem[\protect\citeauthoryear{McGregor and Lee}{McGregor and
  Lee}{2003}]{McGregor03}
{\sc McGregor, J.~P.} {\sc and} {\sc Lee, R.~B.} 2003.
\newblock Architectural techniques for accelerating subword permutations with
  repetitions.
\newblock {\em IEEE Trans. VLSI\/}~{\em 11}, 325--335.

\bibitem[\protect\citeauthoryear{Miller and Sasanian}{Miller and
  Sasanian}{2010}]{Miller10}
{\sc Miller, D.} {\sc and} {\sc Sasanian, Z.} 2010.
\newblock Improving the {NCV} realization of multiple-control {Toffoli} gates.
\newblock {\em Int'l Workshop on Boolean Problems\/}, 37--44.

\bibitem[\protect\citeauthoryear{Miller, Maslov, and Dueck}{Miller
  et~al\mbox{.}}{2003}]{MillerDAC03}
{\sc Miller, D.~M.}, {\sc Maslov, D.}, {\sc and} {\sc Dueck, G.~W.} 2003.
\newblock A transformation based algorithm for reversible logic synthesis.
\newblock {\em Design Autom. Conf.\/}, 318--323.

\bibitem[\protect\citeauthoryear{Miller, Wille, and Drechsler}{Miller
  et~al\mbox{.}}{2010}]{MillerISMVL10}
{\sc Miller, D.~M.}, {\sc Wille, R.}, {\sc and} {\sc Drechsler, R.} 2010.
\newblock Reducing reversible circuit cost by adding lines.
\newblock {\em Int'l Symp. on Multiple-Valued Logic\/}, 217--222.

\bibitem[\protect\citeauthoryear{Mishchenko and Perkowski}{Mishchenko and
  Perkowski}{2002}]{ws:mp}
{\sc Mishchenko, A.} {\sc and} {\sc Perkowski, M.} 2002.
\newblock Logic synthesis of reversible wave cascades.
\newblock {\em Int'l Workshop on Logic Synth.\/}, 197--–202.

\bibitem[\protect\citeauthoryear{Moore and Crutchfield}{Moore and
  Crutchfield}{2000}]{Moore:2000}
{\sc Moore, C.} {\sc and} {\sc Crutchfield, J.~P.} 2000.
\newblock Quantum automata and quantum grammars.
\newblock {\em Theor. Comput. Sci.\/}~{\em 237,\/}~1-2, 275--306.

\bibitem[\protect\citeauthoryear{Morita}{Morita}{2008}]{Morita08}
{\sc Morita, K.} 2008.
\newblock Reversible computing and cellular automata - a survey.
\newblock {\em Theor. Comput. Sci.\/}~{\em 395,\/}~1, 101--131.

\bibitem[\protect\citeauthoryear{{M\"{o}tt\"{o}nen} and
  {Vartiainen}}{{M\"{o}tt\"{o}nen} and {Vartiainen}}{2006}]{Mottonen06}
{\sc {M\"{o}tt\"{o}nen}, M.} {\sc and} {\sc {Vartiainen}, J.~J.} 2006.
\newblock {\em Decompositions of general quantum gates}.
\newblock Ch. 7 in Trends in Quantum Computing Research, NOVA Publishers.

\bibitem[\protect\citeauthoryear{Negrevergne, Mahesh, Ryan, Ditty, Cyr-Racine,
  Power, Boulant, Havel, Cory, and Laflamme}{Negrevergne
  et~al\mbox{.}}{2006}]{Negrevergne06}
{\sc Negrevergne, C.}, {\sc Mahesh, T.~S.}, {\sc Ryan, C.~A.}, {\sc Ditty, M.},
  {\sc Cyr-Racine, F.}, {\sc Power, W.}, {\sc Boulant, N.}, {\sc Havel, T.},
  {\sc Cory, D.~G.}, {\sc and} {\sc Laflamme, R.} 2006.
\newblock Benchmarking quantum control methods on a 12-qubit system.
\newblock {\em Phys. Rev. Lett.\/}~{\em 96,\/}~17.

\bibitem[\protect\citeauthoryear{Nielsen and Chuang}{Nielsen and
  Chuang}{2000}]{Nielsen00}
{\sc Nielsen, M.} {\sc and} {\sc Chuang, I.} 2000.
\newblock {\em Quantum Computation and Quantum Information}.
\newblock Cambridge Univ. Press, Cambridge, UK.

\bibitem[\protect\citeauthoryear{Patel, Markov, and Hayes}{Patel
  et~al\mbox{.}}{2008}]{PatelQIC08}
{\sc Patel, K.~N.}, {\sc Markov, I.~L.}, {\sc and} {\sc Hayes, J.~P.} 2008.
\newblock Optimal synthesis of linear reversible circuits.
\newblock {\em Quant. Inf. Comput.\/}~{\em 8,\/}~3-4, 282--294.

\bibitem[\protect\citeauthoryear{Peres}{Peres}{1985}]{Peres85}
{\sc Peres, A.} 1985.
\newblock Reversible logic and quantum computers.
\newblock {\em Phys. Rev. A\/}~{\em 32}, 3266--3276.

\bibitem[\protect\citeauthoryear{P\'erez-Delgado, Mosca, Cappellaro, and
  Cory}{P\'erez-Delgado et~al\mbox{.}}{2006}]{PhysRevLett.97.100501}
{\sc P\'erez-Delgado, C.~A.}, {\sc Mosca, M.}, {\sc Cappellaro, P.}, {\sc and}
  {\sc Cory, D.~G.} 2006.
\newblock Single spin measurement using cellular automata techniques.
\newblock {\em Phys. Rev. Lett.\/}~{\em 97,\/}~10, 100501.

\bibitem[\protect\citeauthoryear{Polian, Fiehn, Becker, and Hayes}{Polian
  et~al\mbox{.}}{2005}]{Polian:2005}
{\sc Polian, I.}, {\sc Fiehn, T.}, {\sc Becker, B.}, {\sc and} {\sc Hayes,
  J.~P.} 2005.
\newblock A family of logical fault models for reversible circuits.
\newblock {\em Asian Test Symp.\/}, 422--427.

\bibitem[\protect\citeauthoryear{Politi, Matthews, and O'Brien}{Politi
  et~al\mbox{.}}{2009}]{Politi09}
{\sc Politi, A.}, {\sc Matthews, J. C.~F.}, {\sc and} {\sc O'Brien, J.~L.}
  2009.
\newblock {Shor's} quantum factoring algorithm on a photonic chip.
\newblock {\em Science\/}~{\em 325,\/}~5945, 1221.

\bibitem[\protect\citeauthoryear{Prasad, Shende, Markov, Hayes, and
  Patel}{Prasad et~al\mbox{.}}{2006}]{PrasadJETC06}
{\sc Prasad, A.~K.}, {\sc Shende, V.~V.}, {\sc Markov, I.~L.}, {\sc Hayes,
  J.~P.}, {\sc and} {\sc Patel, K.~N.} 2006.
\newblock Data structures and algorithms for simplifying reversible circuits.
\newblock {\em J. Emerg. Technol. Comput. Sys.\/}~{\em 2,\/}~4, 277--293.

\bibitem[\protect\citeauthoryear{Rentergem, Vos, and Keyser}{Rentergem
  et~al\mbox{.}}{2007}]{Rentergem:2007}
{\sc Rentergem, Y.~V.}, {\sc Vos, A.~D.}, {\sc and} {\sc Keyser, K.~D.} 2007.
\newblock Six synthesis methods for reversible logic.
\newblock {\em Open Sys. \& Inf. Dynamics\/}~{\em 14,\/}~1, 91--116.

\bibitem[\protect\citeauthoryear{Saeedi, Arabzadeh, {Saheb Zamani}, and
  Sedighi}{Saeedi et~al\mbox{.}}{2011a}]{SaeediQIC11}
{\sc Saeedi, M.}, {\sc Arabzadeh, M.}, {\sc {Saheb Zamani}, M.}, {\sc and} {\sc
  Sedighi, M.} 2011a.
\newblock Block-based quantum-logic synthesis.
\newblock {\em Quant. Inf. Comput.\/}~{\em 11,\/}~3-4, 0262--0277.

\bibitem[\protect\citeauthoryear{Saeedi, {Saheb Zamani}, and Sedighi}{Saeedi
  et~al\mbox{.}}{2007a}]{SaeediISVLSI07}
{\sc Saeedi, M.}, {\sc {Saheb Zamani}, M.}, {\sc and} {\sc Sedighi, M.} 2007a.
\newblock On the behavior of substitution-based reversible circuit synthesis
  algorithms: investigation and improvement.
\newblock {\em Int'l Symp. on VLSI\/}, 428--436.

\bibitem[\protect\citeauthoryear{Saeedi, {Saheb Zamani}, Sedighi, and
  Sasanian}{Saeedi et~al\mbox{.}}{2010a}]{SaeediJETC10}
{\sc Saeedi, M.}, {\sc {Saheb Zamani}, M.}, {\sc Sedighi, M.}, {\sc and} {\sc
  Sasanian, Z.} 2010a.
\newblock Reversible circuit synthesis using a cycle-based approach.
\newblock {\em J. Emerg. Technol. Comput. Sys.\/}~{\em 6,\/}~4, 13:1--13:26.

\bibitem[\protect\citeauthoryear{Saeedi, Sedighi, and {Saheb Zamani}}{Saeedi
  et~al\mbox{.}}{2007b}]{SaeediICCAD07}
{\sc Saeedi, M.}, {\sc Sedighi, M.}, {\sc and} {\sc {Saheb Zamani}, M.} 2007b.
\newblock A novel synthesis algorithm for reversible circuits.
\newblock {\em Int'l Conf. on Computer-Aided Design\/}, 65--68.

\bibitem[\protect\citeauthoryear{Saeedi, Sedighi, and {Saheb Zamani}}{Saeedi
  et~al\mbox{.}}{2010b}]{SaeediMEJ10}
{\sc Saeedi, M.}, {\sc Sedighi, M.}, {\sc and} {\sc {Saheb Zamani}, M.} 2010b.
\newblock A library-based synthesis methodology for reversible logic.
\newblock {\em Microelectron. J.\/}~{\em 41,\/}~4, 185--194.

\bibitem[\protect\citeauthoryear{Saeedi, Wille, and Drechsler}{Saeedi
  et~al\mbox{.}}{2011b}]{SaeediQIP11}
{\sc Saeedi, M.}, {\sc Wille, R.}, {\sc and} {\sc Drechsler, R.} 2011b.
\newblock Synthesis of quantum circuits for linear nearest neighbor
  architectures.
\newblock {\em Quant. Inf. Proc.\/}~{\em 10,\/}~3, 355--377.

\bibitem[\protect\citeauthoryear{Shende, Bullock, and Markov}{Shende
  et~al\mbox{.}}{2006}]{Shende06}
{\sc Shende, V.}, {\sc Bullock, S.}, {\sc and} {\sc Markov, I.~L.} 2006.
\newblock Synthesis of quantum-logic circuits.
\newblock {\em IEEE Trans. CAD\/}~{\em 25,\/}~6, 1000--1010.

\bibitem[\protect\citeauthoryear{Shende and Markov}{Shende and
  Markov}{2009}]{Shende09}
{\sc Shende, V.~V.} {\sc and} {\sc Markov, I.~L.} 2009.
\newblock On the {CNOT}-cost of {TOFFOLI} gates.
\newblock {\em Quant. Inf. Comput.\/}~{\em 9,\/}~5-6, 461--486.

\bibitem[\protect\citeauthoryear{Shende, Markov, and Bullock}{Shende
  et~al\mbox{.}}{2004}]{shende-2004}
{\sc Shende, V.~V.}, {\sc Markov, I.~L.}, {\sc and} {\sc Bullock, S.~S.} 2004.
\newblock Minimal universal two-qubit quantum circuits.
\newblock {\em Phys. Rev. A\/}~{\em 69}, 062321.

\bibitem[\protect\citeauthoryear{Shende, Prasad, Markov, and Hayes}{Shende
  et~al\mbox{.}}{2003}]{ShendeTCAD03}
{\sc Shende, V.~V.}, {\sc Prasad, A.~K.}, {\sc Markov, I.~L.}, {\sc and} {\sc
  Hayes, J.~P.} 2003.
\newblock Synthesis of reversible logic circuits.
\newblock {\em IEEE Trans. CAD\/}~{\em 22,\/}~6, 710--722.

\bibitem[\protect\citeauthoryear{Shi, Duan, and Vidal}{Shi
  et~al\mbox{.}}{2006}]{Shi06}
{\sc Shi, Y.-Y.}, {\sc Duan, L.-M.}, {\sc and} {\sc Vidal, G.} 2006.
\newblock Classical simulation of quantum many-body systems with a tree tensor
  network.
\newblock {\em Phys. Rev. A\/}~{\em 74,\/}~2, 022320.

\bibitem[\protect\citeauthoryear{Shi and Lee}{Shi and Lee}{2000}]{Shi:2000}
{\sc Shi, Z.} {\sc and} {\sc Lee, R.~B.} 2000.
\newblock Bit permutation instructions for accelerating software cryptography.
\newblock {\em Int'l Conf. on Applic.-Spec. Sys., Architectures, and
  Processors\/}, 138--148.

\bibitem[\protect\citeauthoryear{Skinner, Davenport, and Kane}{Skinner
  et~al\mbox{.}}{2003}]{Skinner03}
{\sc Skinner, A.~J.}, {\sc Davenport, M.~E.}, {\sc and} {\sc Kane, B.~E.} 2003.
\newblock Hydrogenic spin quantum computing in silicon: A digital approach.
\newblock {\em Phys. Rev. Lett.\/}~{\em 90,\/}~8, 087901.

\bibitem[\protect\citeauthoryear{Skoneczny, Van~Rentergem, and
  De~Vos}{Skoneczny et~al\mbox{.}}{2008}]{Skoneczny}
{\sc Skoneczny, M.}, {\sc Van~Rentergem, Y.}, {\sc and} {\sc De~Vos, A.} 2008.
\newblock Reversible fourier transform chip.
\newblock {\em Mixed Design of Integrated Circuits and Sys.\/}, 281--286.

\bibitem[\protect\citeauthoryear{Soeken, Frehse, Wille, and Drechsler}{Soeken
  et~al\mbox{.}}{2010a}]{SFWD:10}
{\sc Soeken, M.}, {\sc Frehse, S.}, {\sc Wille, R.}, {\sc and} {\sc Drechsler,
  R.} 2010a.
\newblock {RevKit}: A toolkit for reversible circuit design.
\newblock {\em {Workshop on Reversible Computation}\/}.
\newblock {RevKit} is available at http://www.revkit.org.

\bibitem[\protect\citeauthoryear{Soeken, Wille, Dueck, and Drechsler}{Soeken
  et~al\mbox{.}}{2010b}]{Soeken10}
{\sc Soeken, M.}, {\sc Wille, R.}, {\sc Dueck, G.~W.}, {\sc and} {\sc
  Drechsler, R.} 2010b.
\newblock Window optimization of reversible and quantum circuits.
\newblock {\em Design \& Diagnostics of Elec. Circ. \& Sys.\/}, 341--345.

\bibitem[\protect\citeauthoryear{Storme, Vos, and Jacobs}{Storme
  et~al\mbox{.}}{1999}]{Storme:jucs_5_5}
{\sc Storme, L.}, {\sc Vos, A.~D.}, {\sc and} {\sc Jacobs, G.} 1999.
\newblock Group theoretical aspects of reversible logic gates.
\newblock {\em J. of Universal Comput. Sci.\/}~{\em 5,\/}~5, 307--321.

\bibitem[\protect\citeauthoryear{Svore, Aho, Cross, Chuang, and Markov}{Svore
  et~al\mbox{.}}{2006}]{Svore:2006}
{\sc Svore, K.~M.}, {\sc Aho, A.~V.}, {\sc Cross, A.~W.}, {\sc Chuang, I.},
  {\sc and} {\sc Markov, I.~L.} 2006.
\newblock A layered software architecture for quantum computing design tools.
\newblock {\em Computer\/}~{\em 39}, 74--83.

\bibitem[\protect\citeauthoryear{Takahashi and Kunihiro}{Takahashi and
  Kunihiro}{2008}]{TakahashiQIC08}
{\sc Takahashi, Y.} {\sc and} {\sc Kunihiro, N.} 2008.
\newblock A fast quantum circuit for addition with few qubits.
\newblock {\em Quant. Inf. Comput.\/}~{\em 8,\/}~6-7, 636--649.

\bibitem[\protect\citeauthoryear{Takahashi, Kunihiro, and Ohta}{Takahashi
  et~al\mbox{.}}{2007}]{Takahashi:2007}
{\sc Takahashi, Y.}, {\sc Kunihiro, N.}, {\sc and} {\sc Ohta, K.} 2007.
\newblock The quantum {Fourier} transform on a linear nearest neighbor
  architecture.
\newblock {\em Quant. Inf. Comput.\/}~{\em 7,\/}~4, 383--391.

\bibitem[\protect\citeauthoryear{Takahashi, Tani, and Kunihiro}{Takahashi
  et~al\mbox{.}}{2010}]{Takahashi:2010}
{\sc Takahashi, Y.}, {\sc Tani, S.}, {\sc and} {\sc Kunihiro, N.} 2010.
\newblock Quantum addition circuits and unbounded fan-out.
\newblock {\em Quant. Inf. Comput.\/}~{\em 10,\/}~9\&10, 872--890.

\bibitem[\protect\citeauthoryear{Toffoli}{Toffoli}{1980}]{Toffoli80}
{\sc Toffoli, T.} 1980.
\newblock Reversible computing.
\newblock Springer, 632.
\newblock Technical Memo MIT/LCS/TM-151, MIT Lab. for Comput. Sci.

\bibitem[\protect\citeauthoryear{{Van Meter} and {Itoh}}{{Van Meter} and
  {Itoh}}{2005}]{Meter05}
{\sc {Van Meter}, R.} {\sc and} {\sc {Itoh}, K.~M.} 2005.
\newblock Fast quantum modular exponentiation.
\newblock {\em Phys. Rev. A\/}~{\em 71,\/}~5, 052320.

\bibitem[\protect\citeauthoryear{{Van Meter} and Oskin}{{Van Meter} and
  Oskin}{2006}]{Meter06}
{\sc {Van Meter}, R.} {\sc and} {\sc Oskin, M.} 2006.
\newblock Architectural implications of quantum computing technologies.
\newblock {\em J. Emerg. Technol. Comput. Sys.\/}~{\em 2,\/}~1, 31--63.

\bibitem[\protect\citeauthoryear{Viamontes, Markov, and Hayes}{Viamontes
  et~al\mbox{.}}{2009}]{Viamontes09}
{\sc Viamontes, G.}, {\sc Markov, I.~L.}, {\sc and} {\sc Hayes, J.~P.} 2009.
\newblock {\em Quantum Circuit Simulation}.
\newblock Springer.

\bibitem[\protect\citeauthoryear{Viamontes, Markov, and Hayes}{Viamontes
  et~al\mbox{.}}{2007}]{Viamontes:2007}
{\sc Viamontes, G.~F.}, {\sc Markov, I.~L.}, {\sc and} {\sc Hayes, J.~P.} 2007.
\newblock Checking equivalence of quantum circuits and states.
\newblock {\em Int'l Conf. on Computer-Aided Design\/}, 69--74.

\bibitem[\protect\citeauthoryear{Visan, Polyakov, Solanki, Arya, Denniston, and
  Cooperman}{Visan et~al\mbox{.}}{2009}]{Visan09}
{\sc Visan, A.~M.}, {\sc Polyakov, A.}, {\sc Solanki, P.~S.}, {\sc Arya, K.},
  {\sc Denniston, T.}, {\sc and} {\sc Cooperman, G.} 2009.
\newblock Temporal debugging using {URDB}.
\newblock {\em CoRR\/}~{\em abs/0910.5046}.

\bibitem[\protect\citeauthoryear{Von~Neumann}{Von~Neumann}{1966}]{Neumann66}
{\sc Von~Neumann, J.} 1966.
\newblock {\em {Theory of Self-Reproducing Automata}}.
\newblock Univ. of Illinois Press, USA.

\bibitem[\protect\citeauthoryear{Wille and Drechsler}{Wille and
  Drechsler}{2009}]{WilleDAC09}
{\sc Wille, R.} {\sc and} {\sc Drechsler, R.} 2009.
\newblock {BDD}-based synthesis of reversible logic for large functions.
\newblock {\em Design Autom. Conf.\/}, 270--275.

\bibitem[\protect\citeauthoryear{Wille and Drechsler}{Wille and
  Drechsler}{2010}]{WilleBook}
{\sc Wille, R.} {\sc and} {\sc Drechsler, R.} 2010.
\newblock {\em Towards a Design Flow for Reversible Logic}.
\newblock Springer.

\bibitem[\protect\citeauthoryear{Wille, Gro{\ss}e, Miller, and Drechsler}{Wille
  et~al\mbox{.}}{2009}]{Wille:2009}
{\sc Wille, R.}, {\sc Gro{\ss}e, D.}, {\sc Miller, D.~M.}, {\sc and} {\sc
  Drechsler, R.} 2009.
\newblock Equivalence checking of reversible circuits.
\newblock {\em Int'l Symp. on Multiple-Valued Logic\/}, 324--330.

\bibitem[\protect\citeauthoryear{Wille, Gro{\ss}e, Teuber, Dueck, and
  Drechsler}{Wille et~al\mbox{.}}{2008a}]{WilleMVL08}
{\sc Wille, R.}, {\sc Gro{\ss}e, D.}, {\sc Teuber, L.}, {\sc Dueck, G.~W.},
  {\sc and} {\sc Drechsler, R.} 2008a.
\newblock {RevLib}: An online resource for reversible functions and reversible
  circuits.
\newblock {\em Int'l Symp. on Multiple-Valued Logic\/}, 220--225.

\bibitem[\protect\citeauthoryear{Wille, Le, Dueck, and Gro{\ss}e}{Wille
  et~al\mbox{.}}{2008b}]{WilleDATE08}
{\sc Wille, R.}, {\sc Le, H.~M.}, {\sc Dueck, G.~W.}, {\sc and} {\sc Gro{\ss}e,
  D.} 2008b.
\newblock Quantified synthesis of reversible logic.
\newblock {\em Design, Autom., and Test Eur.\/}, 1015--1020.

\bibitem[\protect\citeauthoryear{Wille, Offermann, and Drechsler}{Wille
  et~al\mbox{.}}{2010a}]{WOD:2010}
{\sc Wille, R.}, {\sc Offermann, S.}, {\sc and} {\sc Drechsler, R.} 2010a.
\newblock {SyReC: A Programming Language for Synthesis of Reversible Circuits}.
\newblock In {\em Forum on specification \& Design Languages}.

\bibitem[\protect\citeauthoryear{Wille, Soeken, and Drechsler}{Wille
  et~al\mbox{.}}{2010b}]{WilleDAC10}
{\sc Wille, R.}, {\sc Soeken, M.}, {\sc and} {\sc Drechsler, R.} 2010b.
\newblock Reducing the number of lines in reversible circuits.
\newblock {\em Design Autom. Conf.\/}, 647--652.

\bibitem[\protect\citeauthoryear{Yamashita and Markov}{Yamashita and
  Markov}{2010}]{Yamashita10}
{\sc Yamashita, S.} {\sc and} {\sc Markov, I.~L.} 2010.
\newblock Fast equivalence-checking for quantum circuits.
\newblock {\em Quant. Inf. Comput.\/}~{\em 9,\/}~9-10, 721--734.

\bibitem[\protect\citeauthoryear{Yang, Song, Hung, Xie, and Perkowski}{Yang
  et~al\mbox{.}}{2006}]{YangLNCS06}
{\sc Yang, G.}, {\sc Song, X.}, {\sc Hung, W.~N.}, {\sc Xie, F.}, {\sc and}
  {\sc Perkowski, M.~A.} 2006.
\newblock Group theory based synthesis of binary reversible circuits.
\newblock {\em Lec. Notes in Comp. Sci.\/}~{\em 3959/2006}, 365--374.

\bibitem[\protect\citeauthoryear{Yang, Song, Hung, and Perkowski}{Yang
  et~al\mbox{.}}{2008}]{Yang:2008}
{\sc Yang, G.}, {\sc Song, X.}, {\sc Hung, W. N.~N.}, {\sc and} {\sc Perkowski,
  M.~A.} 2008.
\newblock Bi-directional synthesis of 4-bit reversible circuits.
\newblock {\em Comput. J.\/}~{\em 51}, 207--215.

\bibitem[\protect\citeauthoryear{Yokoyama, Axelsen, and Gl\"{u}ck}{Yokoyama
  et~al\mbox{.}}{2008}]{Yokoyama:2008}
{\sc Yokoyama, T.}, {\sc Axelsen, H.~B.}, {\sc and} {\sc Gl\"{u}ck, R.} 2008.
\newblock Principles of a reversible programming language.
\newblock {\em Comput. Frontiers\/}, 43--54.

\bibitem[\protect\citeauthoryear{Zhang, Vala, Sastry, and Whaley}{Zhang
  et~al\mbox{.}}{2003}]{Zhang03}
{\sc Zhang, J.}, {\sc Vala, J.}, {\sc Sastry, S.}, {\sc and} {\sc Whaley,
  K.~B.} 2003.
\newblock Geometric theory of nonlocal two-qubit operations.
\newblock {\em Phys. Rev. A\/}~{\em 67,\/}~4, 042313.

\end{thebibliography}
\end{document}